\documentclass{article}
\newcommand{\be}{\begin{equation}}
\newcommand{\ee}{\end{equation}}
\newcommand{\beqn}{\begin{eqnarray}}
\newcommand{\eeqn}{\end{eqnarray}}
\newcommand{\beqnn}{\begin{eqnarray*}}
\newcommand{\eeqnn}{\end{eqnarray*}}

\def\a{\alpha}
\def\b{\beta}

\def\ep{\epsilon}

\begin{document}

\title{Entanglement of resonantly coupled field modes in cavities with
vibrating boundaries}

\author{M. A. Andreata, A. V. Dodonov and
V. V. Dodonov
\thanks{
on leave from Lebedev Physical Institute and Moscow Institute of Physics and
Technology, Russia}
\\
Departamento de F\'{\i}sica, Universidade Federal de
S\~ao Carlos,\\
Via Washington Luiz, km 235, 13565-905  S\~ao Carlos,  SP,  Brasil}
\date{}
\maketitle


\begin{abstract}
        We study time dependence of various measures of entanglement
(covariance entanglement coefficient, purity entanglement coefficient,
normalized distance coefficient, entropic coefficients)
between resonantly coupled modes of the electromagnetic field
in ideal cavities with oscillating boundaries. Two types of cavities are
considered: a three-dimensional cavity possessing eigenfrequencies
$\omega_3=3\omega_1$, whose wall oscillates at the frequency
$\omega_w=2\omega_1$, and a one-dimensional (Fabry--Perot) cavity with
an equidistant spectrum $\omega_n= n\omega_1$, when the distance between
perfect mirrors oscillates at the frequencies $\omega_1$ and $2\omega_1$.
The behaviour of entanglement measures in these cases turns out
to be completely different, although all three coefficients demonstrate
qualitatively similar time dependences in each case (except for some
specific situations, where the covariance entanglement coefficient,
based on traces of covariance submatrices,
seems to be essentially more sensitive to entanglement than other
measures, which are based on determinants of covariance submatrices).
Different initial states of the field are considered: vacuum,
squeezed vacuum, thermal, Fock, and even/odd coherent states.

\end{abstract}

\vspace{5mm}



\section{Introduction}

During the past decade it was recognized that the concept of
{\em entanglement\/}, introduced by Schr\"odinger in 1935
\cite{Sch-ent,S-cat},
is not only one of the most profound in quantum mechanics
(as was shown in the same year by Einstein, Podolsky and Rosen
in their famous paper \cite{Einstein}, albeit without using
explicitly this word),
 but it is crucial for many promising new applications,
such as quantum cryptography, quantum communication and teleportation,
quantum computing, etc. 
This explain a burst of interest to
various problems connected to this concept observed for the past few years.
One of such problems is a search for
{\em quantitative measures of entanglement}.

In the most cases, the measures based on different kinds of
{\em entropies\/} have been considered
\cite{Bar91,Mann95,Ben96,Ved98,Woot98,DonHor99}.
For example, if the total system, consisting of parts $1$ and $2$,
is decribed by means of the statistical operator $\hat\rho_0$,
then the entanglement measure is frequently expressed in terms of the
total and ``partial'' entropies as the ``index of correlation'' \cite{Bar91}
\be
I_c=S_1+S_2 -S_0,
\qquad S_k=-\mbox{Tr}_k\,\hat\rho_k\,\ln\hat\rho_k,
\label{defIc}
\ee
where the reduced statistical operator is defined as, e.g.,
$\hat\rho_1=\mbox{Tr}_2 \hat\rho_0$.
For the total pure states
formula (\ref{defIc}) is reduced to $I_c=2S_1 =2S_2$.

However, despite of many advantages,
the measures such as (\ref{defIc}) are not very convenient from the
practical point of view: in order to calculate them, one has to diagonalize
the reduced statistical operators, and this is rather difficult problem
in the generic case, especially
for infinite-dimensional Hilbert spaces (corresponding to the so called
``continuous variable systems''), except for a few simple special cases.
Therefore, many authors looked for other measures, which could be
calculated more easily.

In our paper, we consider several families of the simplified measures.
The first one is based on the notions of {\em quantum purity\/}
$\mu=\mbox{Tr}\,\hat\rho^2$ or ``linear entropy'' $S^L=1-\mu$.
Different measures containing these quantities were proposed in
\cite{Ved98,Woot98,DonHor99,3Hor96,Fur98}.
Measures based on the Hilbert--Schmidt distance between the given
state and its ``disentangled'' counterpart were proposed in
\cite{Witte99,MMSZ}.
On the other hand, even simpler (although non-universal)
measures of entanglement of continuous variable quantum
systems, expressed in terms of the cross-covariances
of the quadrature components or the annihilation/creation operators,
have been introduced recently in \cite{JRLR,PLA}.
These measures are discussed in Section 2.

One of numerous possible applications of the entanglement measures
is a compact quantitative characterization of the evolution of coupled
quantum mechanical systems. We began these studies in \cite{JRLR},
where two harmonic oscillators with constant frequencies but with the
most general time-dependent resonance couplings were considered.
The aim of the present paper is to compare different measures in the
case when the field modes in a cavity are entangled due to the motion
of its boundary (the physical reason of entanglement in this case is
the Doppler effect). This case is reduced to the models of two or many
oscillators with {\em time-dependent\/} frequencies and a specific
time-dependent coupling (of the ``coordinate--momentum'' type).

We consider two types of cavities, beginning with
a three-dimensional cavity with accidental degeneracy of
the spectrum (which happens, e.g., in cubical cavities), when only two
modes can occur in resonance with an oscillating wall (Section 3).
A one-dimensional (Fabry-Perot) cavity is considered
in Section 4. In this case {\em all\/} modes are coupled due to
the equidistance
of the (unperturbed) spectrum of the field eigenfrequencies.
It was discovered as far back as in \cite{DK96} that the field evolution
in three- and one-dimensional cavities is qualitatively different.
For example, in the 3D case the number of photons in the resonance
modes grows with time exponentially, whereas in the 1D case this growth
is only linear. It was pointed out in \cite{DK96} that the growth of the
number of photons in the 1D cavity is slowed down due to a strong
intermode interaction, which is equivalent in this case to entanglement.
Now we are able to give a quantitative characterization of such an
entanglement.
The results of our study are discussed in Section 5.

\section{Purity, distance and covariance measures of entanglement}

\subsection{Purity entanglement measure}
By analogy with definition (\ref{defIc}),
the ``linear entropy of entanglement'' can be defined as
\be
{\cal L}= S^L_1+S^L_2 -S^L_0=
1+\mbox{Tr}\,\hat\rho^2-\mbox{Tr}_1\,\hat\rho_1^2
-\mbox{Tr}_2\,\hat\rho_2^2.
\label{defL}
\ee
Such a definition seems reasonable if
the total system is in a {\em pure\/} quantum state. Then
$\mbox{Tr}\,\hat\rho^2=1$ and
$\mbox{Tr}_1\,\hat\rho_1^2 =\mbox{Tr}_2\,\hat\rho_2^2$, so that
${\cal L}= 2S^L_1 =2S^L_2$. As a matter of fact, only this case was
considered in the earlier studies \cite{3Hor96,Fur98},
where measures of entanglement were identified with
the linear entropy of the state of a subsystem
or with some equivalent quantities, such as the purity itself,
the ``participation ratio'' $1/\mbox{Tr}_k\,\hat\rho_k^2$,
or the ``Renyi entropy'' $S^R= -\ln\left(\mbox{Tr}_k\,\hat\rho_k^2\right)$.

However, if the state of the total system is {\em mixed\/}, then definition
(\ref{defL}) leads to some unexpected consequences.
Consider, for example, a generic {\em Gaussian\/} two-mode state described
by means of the
Wigner function (we assume $\hbar=1$ throughout the paper)
\be
  W({\bf q}) =\left|\det\left( {\cal Q}\right)\right|^{-1/2}
  \exp\left
  [-\frac12 \left({\bf q}-\langle{\bf q}\rangle\right)
  {{\cal Q}}^{-1}\left({\bf q}- \langle {\bf q} \rangle \right) \right],
  \qquad
\int W({\bf q}) d{\bf q}/(2\pi)^2 =1,
\label{Wig-Gauss}
\ee
where ${\bf q}=(x_1,p_1,x_2,p_2)$,
and the symmetrical $4\times 4$ covariance matrix ${\cal Q}$ consists of
$2\times2$ blocks
\be
{\cal Q}=\Vert q_{\alpha\beta}\Vert=
\left\Vert
\begin{array}{cc}
{\cal Q}_{11} & {\cal Q}_{12}\\
{\cal Q}_{21} & {\cal Q}_{22}
\end{array}
\right\Vert ,
\qquad
{\cal Q}_{11} =\tilde{\cal Q}_{11}, \quad
{\cal Q}_{22} =\tilde{\cal Q}_{22}, \quad
{\cal Q}_{12} =\tilde{\cal Q}_{21}
\label{defQmatr}
\ee
(a tilde over matrices means matrix transposition).
The symmetrical real covariances are defined as
\be
q_{\alpha\beta}\equiv \frac12\left(\overline{q_{\alpha}q_{\beta}}
+\overline{q_{\beta}q_{\alpha}}\right)
\equiv \widetilde{q_{\alpha}q_{\beta}},
\qquad
\overline{ab} \equiv
\langle \widehat{a}\widehat{b}\rangle -
\langle\widehat{a}\rangle\langle\widehat{b}\rangle
\label{defcov}
\ee
(in other words, a straight line over the product of two observables
means the {\em ordered\/} centralized average value,
whereas a wide tilde means
the {\em symmetrized\/} centralized average value).
Then
\be
\mu \equiv \mbox{Tr}\,\hat\rho^2=
\int [W({\bf q})]^2 d{\bf q}/(2\pi)^2 =
[\det(2{\cal Q})]^{-1/2} \,.
\label{muGauss}
\ee

For factorized (disentangled) states, ${\cal Q}_{12}\equiv 0$, therefore
$\det{\cal Q}=\det{\cal Q}_{11}\det{\cal Q}_{22}$ and
$\mu=\mu_1\mu_2$, which results in the relations
\[
{\cal L}_{fact} = 1 +\mu_1\mu_2 -\mu_1 -\mu_2 =
\left(1-\mu_1\right)\left(1-\mu_2\right).
\]
Consequently, for mixed states ($\mu<1$), one can meet the situation when
${\cal L}>0$ in the absence of any entanglement,
if $\mu_1 \neq 1$ and $\mu_2 \neq 1$.

It seems better to use the difference
$
{\cal L}_*= {\cal L} - {\cal L}_{fact}= \mu- \mu_1\mu_2 $.
But it tends to zero when $\mu\to 0$.
For this reason, we introduce the normalized
{\em purity entanglement coefficient}
\be
\widetilde{\cal L} = 1 -\frac{\mu_1\mu_2}{\mu} \,.
\label{purencof}
\ee
For the {\em Gaussian\/} states (\ref{Wig-Gauss}) it can be expressed as
\be
\widetilde{\cal L} = 1 -\sqrt{\frac{\det{\cal Q}}
{\det{\cal Q}_{11}\det{\cal Q}_{22}}}
= 1-\sqrt{\det\left(E -
{\cal Q}_{12}{\cal Q}_{22}^{-1}{\cal Q}_{21}{\cal Q}_{11}^{-1}\right)}\,.
\label{purenGaus}
\ee
The second equality (where $E$ stands for the unit matrix) is obtained
with the aid of the known formula for the determinant of a block
matrix \cite{Gant}
\[
\det{\cal Q}=
\det\left({\cal Q}_{11} -{\cal Q}_{12}{\cal Q}_{22}^{-1}
{\cal Q}_{21}\right)\det{\cal Q}_{22}.
\]
In particular, for {\em pure\/} composite states ($\mu=1$) we have
\be
\widetilde{\cal L}= 1- \mu_1^2 = 1-\mu_2^2 =\frac14{\cal L}(4-{\cal L}).
\label{tilLpur}
\ee

A measure of entanglement between two coupled modes resembling
(\ref{purenGaus}) was introduced in \cite{PerBer91}
(where it was named ``group correlation coefficient''):
\be
{\cal K}^2 = 1 -\frac{\det{\cal Q}}
{\det{\cal Q}_{11}\det{\cal Q}_{22}}
=\widetilde{\cal L}\left(2-\widetilde{\cal L}\right).
\label{PerBer}
\ee
In principle, the measures (\ref{purenGaus}) and (\ref{PerBer}) can be
used for arbitrary (not only Gaussian) states, although sometimes
they can give zero value even for truly entangled states (if the matrix
of the second-order variances is factorized, but intermode correlations
exist for higher-order moments). Also, instead of (\ref{purencof})
one could use the following extension of formula (\ref{PerBer}) to
arbitrary states:
\be
\widetilde{\cal K}^2 =
1 -\left(\frac{\mu_1\mu_2}{\mu}\right)^2 \,.
\label{PerBermu}
\ee

\subsection{Distance entanglement measure}

Another possibility to characterize entanglement is to use
the Hilbert-Schmidt distance between the given state and different
``disentangled'' states. It was considered, e.g., in
\cite{Ved98,Witte99,MMSZ}
(analogous approach was developed in \cite{Wun00}
to quantify the ``degree of nonclassicality'' of quantum states).
In \cite{MMSZ}, the entanglement measure was defined as
$\mbox{Tr}\left(\hat\rho-\hat\rho_1\otimes\hat\rho_2\right)^2$.
However, we prefer to normalize it by $\mbox{Tr}\hat\rho^2$, in order that
the entanglement measure would not go to zero for highly mixed states.
Thus we shall consider the following quantity:
\be
{\cal Z} =\frac{
\mbox{Tr}\left(\hat\rho-\hat\rho_1\otimes\hat\rho_2\right)^2}
{\mbox{Tr}\hat\rho^2} \equiv
1 + \frac{\mu_1\mu_2}{\mu} -\frac{2}{\mu}
\mbox{Tr}\left(\hat\rho\cdot\left[\hat\rho_1\otimes\hat\rho_2\right]\right).
\label{Z-meas}
\ee

For any states $\hat\rho$ and $\hat{R}$ one has (the normalization
factor corresponds here to the two-mode case)
\be
\mbox{Tr}(\hat\rho\hat{R})=
\int W_{\rho}({\bf q})W_R({\bf q})d{\bf q}/(2\pi)^2\,.
\label{tr12}
\ee
For the {\em Gaussian\/} states (\ref{Wig-Gauss}) the integrals can be
calculated with the aid of the known formula
\be
\int \exp\left(-{\bf q}A{\bf q} +{\bf b}{\bf q}\right)\,d{\bf q} =
\left[\det(A/\pi)\right]^{-1/2}\exp\left(\frac14{\bf b}A^{-1}{\bf b}\right).
\label{Gaussint}
\ee
The inverse matrix ${\cal Q}^{-1}$ can be represented in the block form
with the aid of the Frobenius formula \cite{Gant}
\be
\left\Vert\begin{array}{cc}
{\cal Q}_{11} & {\cal Q}_{12}\\
{\cal Q}_{21} & {\cal Q}_{22} \end{array}
\right\Vert^{-1}  =
\left\Vert\begin{array}{cc}
{\cal Q}_{11}^{-1} +{\cal Q}_{11}^{-1}{\cal Q}_{12} {\cal Q}_{*}^{-1}
{\cal Q}_{21} {\cal Q}_{11}^{-1} &
-{\cal Q}_{11}^{-1}{\cal Q}_{12} {\cal Q}_{*}^{-1}\\
-{\cal Q}_{*}^{-1}{\cal Q}_{21}{\cal Q}_{11}^{-1} & {\cal Q}_{*}^{-1}
\end{array}
\right\Vert,
\label{Frob}
\ee
\[
{\cal Q}_{*}= {\cal Q}_{22}-{\cal Q}_{21}{\cal Q}_{11}^{-1}{\cal Q}_{12}\,.
\]
Taking into account equations (\ref{Wig-Gauss}), (\ref{defQmatr}),
(\ref{Gaussint}), and (\ref{Frob}), one can verify that the Wigner
function of the factorized state
$\hat\rho_1\otimes\hat\rho_2$ is given by formula (\ref{Wig-Gauss})
with the block-diagonal matrix
\[
{\cal Q}_{d}=
\left\Vert
\begin{array}{cc}
{\cal Q}_{11} & 0\\
0 & {\cal Q}_{22}
\end{array}
\right\Vert,
\]
where matrices ${\cal Q}_{11}$ and ${\cal Q}_{22}$ are {\em the same\/}
as in (\ref{defQmatr})
(this is obvious from the physical point of view). Thus we arrive at
the following expression for the ${\cal Z}$-measure
(it is equivalent, except for the normalizing factor $\mu^{-1}$, to
that given in \cite{MMSZ}, but it is written in more simple explicit
form):
\be
{\cal Z} = 1 +\sqrt{\frac{\det{\cal Q}}
{\det{\cal Q}_{11}\det{\cal Q}_{22}}}
-2\sqrt{\frac{\det(2{\cal Q})}
{\det{\cal Q}_{z}}},
\qquad
{\cal Q}_{z} = {\cal Q} + {\cal Q}_{d}=
\left\Vert
\begin{array}{cc}
2{\cal Q}_{11} & {\cal Q}_{12}\\
{\cal Q}_{21} & 2{\cal Q}_{22}
\end{array}
\right\Vert .
\label{ZGaus}
\ee

\subsection{Covariance entanglement measures}
Other measures of entanglement have been introduced
recently in \cite{JRLR,PLA}.
They are expressed directly in terms of the
cross-covariances of the quadrature components or the equivalent
annihilation/creation operators
as follows:
\beqn
{\cal Y}&=&\left[\frac{\mbox{Tr}\left({\cal Q}_{12}
{\cal Q}_{21}\right)}{\mbox{Tr}{\cal Q}_{11}\mbox{Tr}{\cal Q}_{22}}
\right]^{1/2} \\
\label{Ycomp}
&=& \left[ \frac{
|\overline{a_{1}a_{2}^{\dagger}}|^2 + |\overline{a_{1}a_{2}}|^2}
{2\left(\overline{a_{1}^{\dagger}a_{1}}+1/2\right)
\left(\overline{a_{2}^{\dagger}a_{2}}+1/2\right)}
\right]^{1/2}
=  \left[ \frac{
\left(\overline{x_1 x_2}\right)^2
+\left(\overline{p_1 p_2}\right)^2
+\left(\overline{x_1 p_2}\right)^2
+\left(\overline{p_1 x_2}\right)^2
}
{4{\cal E}_1 {\cal E}_2 }
\right]^{1/2},
\label{Y}
\eeqn
\be
\tilde{\cal Y}=\frac{2\sqrt{\mbox{Tr}\left({\cal Q}_{12}
{\cal Q}_{21}\right)}}{\mbox{Tr}{\cal Q}}
\label{def-Ytil}
=
 \frac{\sqrt{2\left(
|\overline{a_{1}a_{2}^{\dagger}}|^2 + |\overline{a_{1}a_{2}}|^2\right)}}
{\overline{a_{1}^{\dagger}a_{1}}+
\overline{a_{2}^{\dagger}a_{2}}+1}
=   \frac{\sqrt{
\left(\overline{x_1 x_2}\right)^2
+\left(\overline{p_1 p_2}\right)^2
+\left(\overline{x_1 p_2}\right)^2
+\left(\overline{p_1 x_2}\right)^2 }
}
{{\cal E}_1+ {\cal E}_2 } , 
\ee
where (we use properly normalized dimensionless quadrature variables)
\be
\hat{a}_k=\left(\hat{x}_k +i\hat{p}_k\right)/\sqrt2, \qquad
{\cal E}_k= \overline{a_{k}^{\dagger}a_{k}} +\frac12
\equiv \frac12\left(
\overline{x_k x_k} +\overline{p_k p_k}\right),
\quad k=1,2.
\label{def-en}
\ee
Since the coefficients (\ref{Ycomp}) and (\ref{def-Ytil}) are expressed
in terms of {\em traces\/}
of products of the off-diagonal blocks of the total covariance
matrix ${\cal Q}$, they are obviously invariant with respect to the
rotations in the phase plane of each subsystem.
(Another invariant quantity, namely the {\em determinant\/}
of the off-diagonal blocks, $\det{\cal Q}_{12}$,
plays an important role for the problem of
{\em separability\/} of continuous variable systems \cite{Duan00}).
It can be shown that $0\le \tilde{\cal Y} \le {\cal Y} <1$.

We would like to emphasize that
the coefficients ${\cal Y}$ and $\tilde{\cal Y}$
are defined for {\em any\/} (not only Gaussian) quantum state.
They are significantly simpler than other entanglement measures
from the point of view of calculations (to calculate traces of matrices
is much more easy than to calculate determinants,
not speaking on calculating eigenvalues of density operators or matrices,
which are necessary to obtain the entropic measures).
A disadvantage of the coefficients ${\cal Y}$ and $\tilde{\cal Y}$ is that
in some cases they are equal to zero even when the state is entangled,
but the second-order moments of quadrature components are equal to zero.
However, this does not happen for Gaussian and many other
important quantum states.

\subsection{Entropic measures for Gaussian states}

In order to demonstrate how simple are expressions given in the preceding
subsections, compared with the ``standard'' entropic measure (\ref{defIc}),
we give
here the formula for the entropy of a generic Gaussian state. It is also
determined by the covariance matrix, but in a more complicated way than
the coefficients considered above. For an arbitrary $N$-mode
Gaussian state the entropy was found in different but equivalent forms
in \cite{167,LukPer89} and recently in \cite{Holev99}. The most simple
expression is \cite{book}
\be
S_N=\sum_{j=1}^{N}\Big[ \left(\kappa_j+1/2\right)\ln
\left(\kappa_j+1/2\right) -
\left(\kappa_j-1/2\right)\ln\left(\kappa_j-1/2\right)\Big],
\label{entrNmod}
\ee
where $\kappa_j\ge 1/2$ ($j=1,\ldots,N$) are $N$ {\em positive\/}
eigenvalues of matrix ${\cal X}$,
which is a ``ratio'' of the symmetric covariance matrix ${\cal Q}$
and antisymmetric commutator matrix:
\be
{\cal X}\equiv {\cal Q}\,\Omega^{-1}, \qquad
{\cal Q}_{jk}=\frac12\langle \overline{ q_j q_k} \!+\!
\overline{ q_k q_j} \rangle,
\qquad
\Omega_{jk}= \langle \hat q_j\hat q_k \!-\! \hat q_k\hat q_j \rangle.
\label{defXQOm}
\ee
One can easily verify that if $\kappa$ is an eigenvalue of ${\cal X}$,
then $-\kappa$ is another eigenvalue. Also, it can be shown that all
eigenvalues of ${\cal X}$ are real.
It is worth emphasizing that formula (\ref{entrNmod}) is valid for
arbitrary sets of operators with $c$-number commutators (canonical
coordinates and momenta, ``annihilation'' and ``creation'' operators,
kinetic momenta and relative coordinates for particles moving in
homogeneous magnetic fields, etc.).

In the one-mode case,
the eigenvalues of matrix ${\cal X}$  are equal to
$\pm \kappa$ (and ${\cal X}^2 \!=\! \kappa^2 E_2$), where
\be
 \kappa=\hbar^{-1}\sqrt{\Delta}\,, \qquad
\Delta  \equiv
\overline{xx}\;\overline{pp}
-\left(\widetilde{xp}\right)^2 \ge \hbar^2/4
= \det{\cal Q}\,.
\label{def-kap}
\ee
(The last inequality is the Schr\"odinger--Robertson uncertainty relation
\cite{SR,DKM80,183}.)
In this case, different expressions equivalent to formula
(\ref{entrNmod}) were found in \cite{167,Agentr,Per86a}.

Calculating the characteristic polynomial of the $4\times4$ matrix
${\cal X}$ in the two-mode case, one arrives at the {\em biquadratic\/}
equation
(for $\hbar=1$)
\cite{LukPer89}
\be
\kappa^4 -{\cal D}_2 \kappa^2 + {\cal D}_0 =0,
\label{chareq}
\ee
where coefficients ${\cal D}_2$ and ${\cal D}_0$ are nothing but
{\em quantum universal invariants\/}, i.e., functions which are
invariant
with respect to arbitrary linear canonical (preserving commutation relations)
transformations \cite{univ}:
\be
{\cal D}_2 = \Delta_1 + \Delta_2 +
2\left(\overline{x_1x_2}\;\overline{p_1p_2}
-\overline{p_1x_2}\;\overline{p_2x_1}\right),
\label{univD2}
\ee
\beqn
D_0^{(2)} &=& \det{\cal Q} =
\left(\overline{p_{1}^2}\;\overline{p_{2}^{2}}
-\overline{p_{1}p_{2}}^{2}\right)
\left(\overline{x_{1}^2}\;\overline{x_{2}^{2}}
-\overline{x_{1}x_{2}}^{2}\right)
+\left(\widetilde{x_1p_1}\;\widetilde{x_2p_2}
- \overline{x_1p_2}\;\overline{x_2p_1}\right)^2
\nonumber\\&&
-\overline{x_2^2}\;\overline{p_1^2}\left(\overline{x_1p_2}\right)^2
-\overline{x_1^2}\;\overline{p_2^2}\left(\overline{x_2p_1}\right)^2
-\overline{x_2^2}\;\overline{p_2^2}
\left(\widetilde{x_1p_1} \right)^2
-\overline{x_1^2}\;\overline{p_1^2}
\left(\widetilde{x_2p_2} \right)^2
\nonumber\\&&
+2\,\overline{x_1x_2}\,
\left[\overline{p_1^2}\;\overline{x_1p_2}
\;\widetilde{x_2p_2}
+\overline{p_2^2}\;\overline{x_2p_1}
\;\widetilde{x_1p_1}  \right]
+2\,\overline{p_1p_2}\,
\left[\overline{x_2^2}\;\overline{x_1p_2}\;\widetilde{x_1p_1}
+\overline{x_1^2}\;\overline{x_2p_1}\;\widetilde{x_2p_2}\right]
\nonumber\\&&
-2\,\overline{x_1x_2}\;\overline{p_1p_2}\left(\widetilde{x_1p_1}\;
\widetilde{x_2p_2}
+\overline{x_1p_2}\;\overline{x_2p_1}\right)
\,.\label{univD0}
\end{eqnarray}
The symbol $\Delta_k$, obviously, means the combination defined by
(\ref{def-kap}) and related to the $k$th mode.

Positive solutions of Eq. (\ref{chareq}) read
\be
\kappa_{1,2} =\frac12 \left[\sqrt{{\cal D}_2 + 2\sqrt{{\cal D}_0}} \pm
\sqrt{{\cal D}_2 - 2\sqrt{{\cal D}_0}} \right].
\label{solchareq}
\ee
The reality of $\kappa_{1,2}$ is ensured by the inequalities
\be
{\cal D}_2  \ge 2\sqrt{{\cal D}_0}
\ge \frac{\hbar^2}{2},
\label{ineq}
\ee
which can be considered as generalized uncertainty relations for two-mode
systems (for systematic studies of such generalizations see, e.g.,
\cite{183,Trif}).

Formulae (\ref{entrNmod}), (\ref{def-kap})  and (\ref{solchareq}) permit
us to express the entropic index of correlation (\ref{defIc})
analytically in terms of the covariances of quadrature components
for arbitrary Gaussian states. However, the corresponding expression
is very cumbersome, and it is much more complicated than any other
entanglement measure discussed in the preceding subsections.
In the next sections we compare the behaviour of different entanglement
measures for various concrete physical models.

\section{A three-dimensional cavity with a vibrating wall and two resonantly
coupled modes}

Classical and quantum phenomena in cavities with moving boundaries attracted
attention of many researchers for a long time (see review \cite{review}).
Especially popular this topic
became in the last decade, being known now under the names {\it %
nonstationary Casimir effect\/} \cite{DKM89},
{\it dynamical Casimir effect\/} \cite{Sch},
or {\it mirror (motion) induced radiation\/} \cite{BE,Lamb}.
One of several theoretical results obtained in the last years was the
prediction of the exponential growth of the energy of the field under the
resonance conditions, when the wall performs vibrations at the frequency
which is a multiple of the unperturbed field eigenfrequency
\cite{DK96,Lamb,Law}.

A unified description of the field inside an ideal cavity with moving
boundaries can be achieved in the frameworks of
the Hamiltonian approach proposed by Law \cite{Law} and
developed in \cite{Plun} (for other references see \cite{review},
and for the most recent publications see
\cite{Plun02,SaiHy02,MaMa02}).
Consider a scalar massless field $\Phi({\bf r},t)$, satisfying
the wave equation $\Phi_{tt}=\nabla^2\Phi$ inside the cavity and the
Dirichlet boundary condition $\Phi=0$ on the boundary
(we assume $c=\hbar=1$).
We assume that we know the complete orthonormalized set of eigenfunctions
(and eigenfrequencies) of the Laplace equation
$\nabla^2 f_{\a}({\bf r}) +\omega_{\a}^2 f_{\a}({\bf r})=0$
in the case of stationary cavity.
Now suppose that a part of the boundary is a plane surface moving
according to a {\em prescribed\/} law of motion $L(t)$ (for the most recent
study of the case when $L(t)$ is a {\em dynamical variable\/} due to
the back reaction of the field see \cite{Cole01}).
Expanding the field $\Phi({\bf r},t)$ over ``instantaneous''
eigenfunctions $f_{\a}({\bf r};L(t))$,
\be
\Phi({\bf r},t)=\sum_{\a} q_{\a}(t) f_{\a}({\bf r};L(t)),
\label{expansion}
\ee
we satisfy automatically the boundary conditions. Then the dynamics of the
field is described completely by the dynamics of the generalized
coordinates $q_{\a}(t)$, which, in turn,
can be derived from the
{\em time-dependent Hamiltonian\/} \cite{Plun}
\be
H(t) = \frac12\sum_{\a}\left[p_{\a}^2 +\omega_{\a}^2(L(t)) q_{\a}^2\right]
+\frac{\dot{L}(t)}{L(t)}\sum_{\a \neq \beta} p_{\a} m_{\a\beta}q_{\beta}
\label{genHam}
\ee
with antisymmetrical time-independent coefficients
\be
m_{\a\beta}= - m_{\beta\a} =L \int dV
\frac{\partial f_{\a}({\bf r};L)}{\partial L}f_{\beta}({\bf r};L).
\label{defmab}
\ee
For example, in the case of a rectangular three-dimensional cavity
with dimensions $L_x,L_y,L_z$, the eigenmodes are well known
products of sine functions like $\sin\left(\pi k_x x/L_x\right)$
(or sine and cosine functions in the case of electromagnetic field),
labeled by three
natural numbers $k_x,k_y,k_z$, whereas unperturbed eigenfrequencies are given
by the formula
\be
\omega_{k_x,k_y,k_z}=\pi\sqrt{\left(\frac{k_x}{L_x}\right)^2
+\left(\frac{k_y}{L_y}\right)^2 +\left(\frac{k_z}{L_z}\right)^2 }\,.
\label{spectr}
\ee
If one surface of the parallelepiped,
perpendicular to the $x$-axis, moves in the $x$-direction
(so that the $L_x$-dimension of the cavity is a function of time), then
\cite{Maz01}
\be
m_{{\bf k}{\bf j}} = (-1)^{k_x+j_x} \frac{2 k_x j_x}{j_x^2 - k_x^2}
\delta_{k_y j_y}\delta_{k_z j_z}.
\label{m3D}
\ee
(In the case of electromagnetic field, one should take into account
polarizations
of the modes, i.e., that $f_{\a}$ and $f_{\b}$ in Eq. (\ref{defmab})
are vector functions, whose directions are perpendicular, respectively,
to the vectors
$(k_x,k_y,k_z)$ and $(j_x,j_y,j_z)$. But one can always
choose two modes with coinciding polarizations, directed along the
perpendicular
to the plane formed by these two vectors. Then all formulae are the same
as in the scalar case.)


We are interested in the case when one of the cavity's walls performs
small oscillations with the frequency $\Omega$ close to the double frequency
of some unperturbed mode $\omega_1^{(0)} \equiv 1$ (i.e., we normalize all
frequencies by $\omega_1^{(0)}$), so that the time-dependent frequency
$\omega_1(t)$ reads
\be
\omega_1(t) = 1 + 2\ep \cos(2\overline{\omega}t),
\quad \overline{\omega}=1 +\delta,
\label{Lt}
\ee
where we assume that $|\delta|\ll 1$ and $|\ep|\ll 1$.
Also we suppose that the unperturbed field frequency spectrum includes
the frequency $\omega_3^{(0)} =3 +\Delta$ with $|\Delta|\ll 1$,
but it does not
contain frequencies close to $5\omega_1^{(0)}$.
A possibility of such a situation was pointed out in \cite{Maz01}.
An example is a cubic cavity with the pair of modes $\{111\}$ and $\{511\}$.
Another example is the pair of modes $\{110\}$ and $\{510\}$ in the
rectangular cavity with $L_x=\sqrt{2}\,L_y$ (in this case, the common
direction of polarization is along the $z$-axis).
Then we have two
resonantly interacting modes, and it is sufficient to consider only the
part of the total Hamiltonian (\ref{genHam}) related to these modes
\cite{Sasha}
(hereafter we use the symbols $x_k$, $p_j$ instead of $q_k$, $p_j$ for
the quadrature components of the field, whereas the letter $x$ without
indices will mean the usual space coordinate inside the cavity):
\be
H_{13} =\frac12\left(p_1^2 + p_3^2\right) + \frac12
\left[1 + 4\ep \cos(2\overline{\omega}t)\right] x_1^2
+\frac12 \left[9+6\Delta +\tilde{\ep} \cos(2\overline{\omega}t)\right]x_3^2
+3\mu\ep\sin(2\overline{\omega}t)\left(p_1 x_3 -p_3 x_1\right).
\label{H2}
\ee
The constant parameter $\mu$
is proportional to the coefficient $m_{12}$ in (\ref{genHam}).
For the rectangular cavity, $\mu=j_x/(12k_x)$
if the modes $\{k_x,m,n\}$ and $\{j_x,m,n\}$ are in resonance.
Writing (\ref{H2}) we have neglected the
second order terms with respect to $\ep$ and $\Delta$.
Parameter $\tilde{\ep}$
has the same order of magnitude as $\ep$,
but it does not affect the solution in the
zeroth order approximation \cite{Sasha}.

Hamiltonian (\ref{H2}) results in the following differential
equations for the generalized coordinates $x_1$ and $x_3$
(we neglect corrections of the second order):
\be
\ddot{x}_1 = -\left[1 + 4\ep \cos(2\overline{\omega}t)\right] x_1
+ 24\mu\ep\left[ \cos(2\overline{\omega}t) x_3
+ \sin(2\overline{\omega}t) \dot{x}_3\right],
\label{ddotx1}
\ee
\be
\ddot{x}_3 =
- \left[9+6\Delta +\tilde{\ep} \cos(2\overline{\omega}t)\right] x_3
- 24\mu\ep\left[ \cos(2\overline{\omega}t) x_1
+ \sin(2\overline{\omega}t) \dot{x}_1\right].
\label{ddotx2}
\ee
These equations have been solved,
using the method of slowly varying amplitudes,
in \cite{Sasha}. We consider here two special cases.

\subsection{Exact (symmetric) resonance} \label{symres}

In the case of exact resonance, $\delta=\Delta=0$,
the solutions of Eqs. (\ref{ddotx1}) and (\ref{ddotx2}) read
\beqn
x_1(t) &=& x_1(0)\left[C_{1}^{-}\,\cos(\rho\tau)
+S_{1}^{-}\,\frac{\sin(\rho\tau)}{\rho}\right]
- p_1(0)\left[S_{1}^{-}\,\cos(\rho\tau)
+C_{1}^{-}\,\frac{\sin(\rho\tau)}{\rho}\right]
\nonumber\\&&
+8\mu\frac{\sin(\rho\tau)}{\rho}
\left[3S_{1}^{-}\, x_3(0) +C_{1}^{-}\, p_3(0)\right],
\label{x1t}
\eeqn
\beqn
x_3(t) &=& x_3(0)\left[C_{3}^{+}\,\cos(\rho\tau)
-S_{3}^{+}\,\frac{\sin(\rho\tau)}{\rho}\right]
+\frac13 p_3(0)\left[S_{3}^{+}\,\cos(\rho\tau)
-C_{3}^{+}\,\frac{\sin(\rho\tau)}{\rho}\right]
\nonumber\\&&
-8\mu\frac{\sin(\rho\tau)}{\rho}
\left[S_{3}^{+}\, x_1(0) -C_{3}^{+}\, p_1(0)\right],
\label{x2t}
\eeqn
\beqn
p_1(t) &=& -x_1(0)\left[S_{1}^{+}\,\cos(\rho\tau)
+C_{1}^{+}\,\frac{\sin(\rho\tau)}{\rho}\right]
+ p_1(0)\left[C_{1}^{+}\,\cos(\rho\tau)
+S_{1}^{+}\,\frac{\sin(\rho\tau)}{\rho}\right]
\nonumber\\&&
-8\mu\frac{\sin(\rho\tau)}{\rho}
\left[3C_{1}^{+}\, x_3(0) +S_{1}^{+}\, p_3(0)\right],
\label{p1t}
\eeqn
\beqn
p_3(t) &=& 3x_3(0)\left[S_{3}^{-}\,\cos(\rho\tau)
-C_{3}^{-}\,\frac{\sin(\rho\tau)}{\rho}\right]
+ p_3(0)\left[C_{3}^{-}\,\cos(\rho\tau)
-S_{3}^{-}\,\frac{\sin(\rho\tau)}{\rho}\right]
\nonumber\\&&
-24\mu\frac{\sin(\rho\tau)}{\rho}
\left[C_{3}^{-}\, x_1(0) -S_{3}^{-}\, p_1(0)\right],
\label{p2t}
\eeqn
where
\be
C_{k}^{\pm}(\tau;t) = \cosh\tau \cos(k\overline{\omega}t)
\pm \sinh\tau \sin(k\overline{\omega}t),  \qquad
S_{k}^{\pm}(\tau;t) = \sinh\tau \cos(k\overline{\omega}t)
\pm \cosh\tau \sin(k\overline{\omega}t),
\label{defSpm}
\ee
\be
\tau \equiv \frac12 \ep t, \qquad \rho=\sqrt{2\nu-1},
\qquad \nu \equiv 96\mu^2 .
\label{def-nu}
\ee
The arguments $t$ (``fast time'') and $\tau$ (``slow time'')
of the functions $C_{k}^{\pm}(\tau;t)$ and $S_{k}^{\pm}(\tau;t)$
can be considered
as independent variables. Then the following relations hold:
\be
\frac{\partial C_{k}^{\pm}}{\partial t}= \pm k S_{k}^{\mp}, \qquad
\frac{\partial S_{k}^{\pm}}{\partial t}= \pm k C_{k}^{\mp}.
\label{difrul}
\ee
For the modes $\{111\}$ and $\{511\}$ of the cubical cavity
or $\{110\}$ and $\{510\}$ of the rectangular cavity with
$L_x=\sqrt{2}\,L_y$
we have $\nu=50/3$. Due to this explicit example,
we assume that parameter $\nu$ is large: $\nu \gg 1 $.

Symbols $x_k$ and $p_k$ in equations (\ref{x1t})-(\ref{p2t})
can be considered
both as classical variables and quantum operators in the Heisenberg
picture, due to the linearity of the problem (or due to the quadratic
nature of Hamiltonian (\ref{genHam})). Using equations
(\ref{x1t})-(\ref{p2t}),
one can calculate mean values of squares and
products of canonical variables (operators) at any moment of time,
provided such mean values were known at the initial moment $t=0$.
We confine ourselves to the case when initially the field
modes were in thermal states with the mean photon numbers
$(\theta_1-1)/2$ and $(\theta_3-1)/2$, where $\theta_k=\coth(k\beta_k/2)$,
$\beta_k$ being inverse absolute temperature in dimensionless units.
In the natural case of equal initial temperatures of the modes,
the following relations hold:
\be
\theta_{31}\equiv \frac{\theta_{3}}{\theta_{1}} =\theta_{13}^{-1}
=\frac{\theta_{1}^2 +3}{3\theta_{1}^2 +1}, \qquad
1\ge \theta_{31} \ge \frac13.
\label{thetratio}
\ee
The {\em normalized\/} mean energies in each mode,
${\cal E}_k =\langle p_k^2 + \omega_k^2 x_k^2\rangle/(2\omega_k)$
(namely these quantities are used in the definitions of the covariance
entanglement coefficients (\ref{Y}) and (\ref{def-Ytil})),
depend on time as follows \cite{Sasha},
\be
{\cal E}_1 = \frac{\theta_1}{2}\Bigg\{
\cosh(2\tau)\left[
\frac{\sin^2(\rho\tau)}{\rho^2}
\left(1 +2\nu\,\theta_{31}\right)
+\cos^2(\rho\tau)\right]
+ \sinh(2\tau) \frac{\sin(2\rho\tau)}{\rho}\Bigg\},
\label{E1t}
\ee
\be
{\cal E}_3 = \frac{\theta_3}{2}\Bigg\{
\cosh(2\tau)\left[
\frac{\sin^2(\rho\tau)}{\rho^2}
\left(1 +2\nu\,\theta_{13}\right)
+\cos^2(\rho\tau)\right]
- \sinh(2\tau) \frac{\sin(2\rho\tau)}{\rho}\Bigg\}.
\label{E2t}
\ee

Calculating the covariance entanglement coefficient, one should use,
instead of variables $x_k$ and $p_k$, the {\em normalized\/} variables
$\tilde{x}_k=\sqrt{\omega_k}\,x_k$ and $\tilde{p}_k=p_k/\sqrt{\omega_k}$
(in our case $\omega_k\equiv k$): see equation (\ref{def-en}).
After some algebra we have obtained the following expressions:
\be
{\cal Y}=\sqrt{\frac{{\cal F}}{4{\cal E}_1{\cal E}_3}}, \qquad
\tilde{\cal Y}=\frac{\sqrt{{\cal F}}}{{\cal E}_1+{\cal E}_3},
\label{Y13}
\ee
\beqn
{\cal F} &=& \frac{\nu}{2\nu-1}\sin^2(\rho\tau)\left\{
\cosh(4\tau)\left[\cos^2(\rho\tau)\left(\theta_1-\theta_3\right)^2
+\frac{\sin^2(\rho\tau)}{\rho^2}\left(\theta_1+\theta_3\right)^2\right]
\right.\nonumber \\ &&\left.
+\frac{\sin(2\rho\tau)}{\rho}\sinh(4\tau)\left(\theta_1^2-\theta_3^2\right)
\right\}.
\label{def-F}
\eeqn
The determinants of the covariance matrices for each mode have been
calculated in \cite{Sasha}.
For the first mode,
\be
\det{\cal Q}_{11}= \frac14\theta_1^2 g_1^2\,, \qquad
g_1^2 = \cos^4(\rho\tau) +
\sin^2(2\rho\tau)\frac{2\nu\,\theta_{31}-1}{2(2\nu-1)}
+ \sin^4(\rho\tau)\left(\frac{2\nu\,\theta_{31}+1}{2\nu-1}
\right)^2.
\label{Dtau}
\ee
For another excited mode
one should interchange indices $1$ and $3$ in (\ref{Dtau}).
Since the evolution of the total system is unitary in the case discussed,
the total determinant does not depend on time:
$\det{\cal Q}=\theta_1^2 \theta_3^2/16$.
Therefore the {\em purity entanglement coefficient\/} (\ref{purenGaus})
has the form
\be
\widetilde{\cal L}= 1 - \left(g_1g_3\right)^{-1}.
\label{Ltil13}
\ee
Eqs. (\ref{entrNmod}) and (\ref{def-kap}) lead to the following
explicit formula for the entropic entanglement measure
$I_c$ (\ref{defIc}):
\be
I_c = \frac12 \sum_{i=1,3}\left[
\left(\theta_i g_i \!+\!1 \right)\ln\left(\theta_i g_i \!+\!1 \right)
-\left(\theta_i g_i \!-\!1 \right)\ln\left(\theta_i g_i \!-\!1 \right)
-\left(\theta_i  \!+\!1 \right)\ln\left(\theta_i  \!+\!1 \right)
+\left(\theta_i  \!-\!1 \right)\ln\left(\theta_i  \!-\!1 \right) \right].
\label{entr13}
\ee
We see that despite the exponential (although non-monotonous in the
high-temperature case $\theta_k \gg 1$ \cite{Sasha})
growth of energy of each mode,
{\em all\/} entanglement coefficients exhibit strong (quasi)periodic
oscillations as functions of the ``slow time'' $\tau$, going to zero
when $\rho\tau=n\pi$.

In the simplest case of the initial
vacuum states of each mode ($\theta_1 =\theta_3= 1$) we have
\be
{\cal F}=\frac{4\nu}{(2\nu-1)^2} \sin^4(\rho\tau)\cosh(4\tau),
\label{Fvac}
\ee
\be
g_1^2=g_3^2 = g_0^2 \equiv 1+ \frac{8\nu}{(2\nu-1)^2}\sin^4(\rho\tau),
\label{g0sym}
\ee
\be
\widetilde{\cal L} = \frac{8\nu\sin^4(\rho\tau)}
{(2\nu-1)^2 + 8\nu\sin^4(\rho\tau)} \approx
\frac{2}{\nu} \sin^4(\rho\tau),
\label{Ltilvac}
\ee
\be
I_c = \left(g_0 \!+\!1 \right)\ln\left(g_0 \!+\!1 \right)
-\left(g_0 \!-\!1 \right)\ln\left(g_0 \!-\!1 \right) -2\ln2
\approx \frac{\sin^{4}(\rho\tau)}{\nu}
\ln\left(\frac{2e\nu}{\sin^{4}(\rho\tau)}\right).
\label{Icsymlow}
\ee
The approximate equalities in (\ref{Ltilvac}) and (\ref{Icsymlow}) hold
for $\nu\gg 1$.
Under this condition,
${\cal E}_1\approx {\cal E}_3\approx \frac12\cosh(2\tau)$,
so that for $\tau>1$ we obtain
\[
{\cal Y} \approx \sqrt{\frac2{\nu}}\,\sin^2(\rho\tau)
\approx\sqrt{\widetilde{\cal L}}.
\]
The evolution of functions $\widetilde{\cal L}(\tau)$ and ${\cal Y}^2(\tau)$
for the initial vacuum case is shown in Figure \ref{figurYLsym}.

For high-temperature initial states ($\theta_{1,3}\gg 1$),
the entanglement coefficients
do not depend on the parameter $\nu$ (if $\nu\gg 1$)
for almost all instants of time, more precisely, under the condition
$|\cos(\rho\tau)|\gg \rho^{-1}\sim \nu^{-1/2}$:
\be
\widetilde{\cal L} \approx \frac{\sin^2(2\rho\tau)
\left(\theta_{31}+\theta_{13}-2\right)}
{4 + \sin^2(2\rho\tau) \left(\theta_{31}+\theta_{13}-2\right)}
\approx {\cal Y}^2 ,
\label{Ltilhight}
\ee
\be
I_c = \ln\left(g_1g_3\right) \approx
\ln\left[1 +\frac14
\sin^2(2\rho\tau) \left(\theta_{31}+\theta_{13}-2\right)\right].
\label{Icsymhigh}
\ee
(The last approximate equality in (\ref{Ltilhight}) holds for $\tau >1$.
The simple formula for $I_c$ is obtained in the limit case
$\theta_{1,3}\to \infty$;
there are some corrections of the order of $\theta_{1,3}^{-1}$
for finite initial mean numbers of photons.)
For the maximal possible value of the coefficient $\theta_{13}=3$
(in the case of true initial thermal equilibrium),
the maximum values of the expressions (\ref{Ltilhight}) and
(\ref{Icsymhigh})
(which are achieved when $\sin^2(2\rho\tau)=1$) are equal to
$\tilde{\cal L}_{max}=1/4$ and $I_c^{(max)} =\ln(4/3)\approx 1/3$.
Note that in the limit high-temperature case,
the purity entanglement coefficient
{\em coincides identically\/} with one of possible forms of the
{\em ``compact entropy''\/}
(another compact parameter, $\tanh(I_c)$, was introduced in \cite{JRLR})
\be
{\cal J}_c =1-\exp\left(-I_c\right).
\label{defJcnew}
\ee
Figure \ref{figIsym} shows the evolution of entropic entanglement measure
$I_c(\tau)$ for vacuum and high-temperature initial states.
Note that one has $\theta_1 \approx 140$, if $L_0=1\,$cm and $T=300\,$K.
For this value of $\theta_1$, the plot of the compact entropy
${\cal J}_c(\tau)$ becomes indistinguishable from the plot of the
purity entanglement coefficient $\widetilde{\cal L}(\tau)$.
The functions $\widetilde{\cal L}(\tau)$ and
${\cal Y}^2(\tau)$ are compared in Figure \ref{figurYLJsym}.
We see that two functions are very close in some intervals, although
their maxima are different (because the value $\rho\approx 5.7$ is not
very large for the chosen parameter $\nu=50/3$).

In the high-temperature case, intermediate nonzero minima
of the entanglement coefficients
(besides exact zero minima at the instants $\tau_n = n\pi/\rho$)
are observed at the moments of ``slow time''
when the modes approximately exchange their purities \cite{Sasha}.
The positions of these additional minima for $\tilde{\cal L}$ and $I_c$
are determined by the condition $\cos(\rho\tau)= 0$, so that
\be
\widetilde{\cal L}_{min} = \frac{2\nu \left(\theta_{31}+\theta_{13}+2\right)}
{4 \nu^2 +1 + 2\nu \left(\theta_{31}+\theta_{13}\right)}, \qquad
I_c^{(min)} =\ln\left[1+
\frac{2\nu \left(\theta_{31}+\theta_{13}+2\right)}{(2\nu-1)^2}\right].
\label{Ltilmin}
\ee
For $\nu \gg 1$ we have
\be
\widetilde{\cal L}_{min} \approx I_c^{(min)} \approx
\frac{\theta_{31}+\theta_{13}+2}{2\nu}.
\label{minhighLI}
\ee
On the other hand, the intermediate minima of ${\cal Y}$ are much smaller.
Indeed, the minimum of the expression inside figure brackets in Eq.
(\ref{def-F}) is achieved for (neglecting corrections of the order of
$\rho^{-3}$)
\[
\tan(2\rho\tau) = \frac2{\rho}\tanh(4\tau)
\frac{\theta_1+\theta_3}{\theta_1-\theta_3}\,.
\]
At this moment of time we obtain
\[
{\cal F} \approx \frac{\left(\theta_{1}+\theta_{3}\right)^2}
{4\nu \cosh(4\tau)}\,, \qquad
4{\cal E}_1 {\cal E}_3 \approx \theta_{1}\theta_{3}\cosh^2(2\tau),
\]
so that for $\tau >1$,
\[
{\cal Y} \approx e^{-4\tau}\sqrt{\frac2{\nu}\left(
\theta_{31}+\theta_{13}+2\right)} \equiv {\cal Y}_*
\approx 2 e^{-4\tau} \sqrt{\widetilde{\cal L}_{min}}\,,
\]
and it is clear that the intermediate minimum of ${\cal Y}$ does not
exceed the value ${\cal Y}_*$.

Therefore, we arrive at rather paradoxical situation, especially
for realistic values of parameters $\nu$ and $\theta_{1,3}$.
According to Figure \ref{figurYLJsym},
the intermediate mimimum value of $\widetilde{\cal L}$-coefficient
in the high-temperature case is only twice less than the maximal value.
Moreover, this high-temperature intermediate mimimum value
is {\em bigger\/} than the maximum value in the vacuum case
(see Figure \ref{figurYLsym}).
Thus, the $\widetilde{\cal L}$-coefficient tells us that for
$\cos(\rho\tau)= 0$,
two modes are ``more entangled'' in the high-temperature case than in
the case of initial vacuum state (or at least
have the same order of entanglement, according to the $I_c$-coefficient
in Figure \ref{figIsym}),
whereas the covariance entanglement coefficient ${\cal Y}$
shows that two modes become practically disentangled at this instant of time.

The resolution of this ``paradox'' is as follows.
According to Eqs. (\ref{Y}), (\ref{Y13}) and (\ref{def-F}),
the function ${\cal F}$ gives the upper limit for squares of
any elements of the ``off-diagonal'' block
${\cal Q}_{12}$ of the covariance matrix ${\cal Q}$ (\ref{defQmatr}),
whereas functions ${\cal E}_k$ give the bounds for the elements
of ``diagonal'' blocks ${\cal Q}_{kk}$. This happens because
${\cal F}$ and ${\cal E}_k$ are based on {\em traces\/} of the covariance
submatrices. Therefore, if ${\cal Y}\to 0$,
this means that {\em all\/} elements
of matrix ${\cal Q}_{12}$ responsible for the intermode
correlations (at least for the Gaussian states considered in this section)
become negligible in comparison with the {\em variances\/}
$\overline{x_kx_k}$ and $\overline{p_kp_k}$
of the quadrature
components. From the physical point of view, it is equivalent to
disappearance of
correlations between the two subsystems, i.e., their disentanglement.

On the other hand, the coefficients $\tilde{\cal L}$ and $I_c$
are based on {\em determinants\/} of the covariance submatrices.
But it is well known that the determinant
of a matrix can be quite small even if all elements of the matrix are big,
and this is the reason of the qualitative difference in the
behaviour of the ``covariance'' and ``entropic'' entanglement coefficients.
This is clearly seen from the last expression in Eq. (\ref{purenGaus}),
which shows that the value of the purity entanglement coefficient
$\widetilde{\cal L}$ depends of the matrix
$R={\cal Q}_{12}{\cal Q}_{22}^{-1}{\cal Q}_{21}{\cal Q}_{11}^{-1}$.
Using easily verified formula
$\det(E+\alpha)\approx \mbox{Tr}\,\alpha$, which holds provided all
elements of matrix $\alpha$ are small with respect to unity,
we can simplify formula (\ref{purenGaus}) in the case of small entanglement
as follows:
\be
\widetilde{\cal L} \approx \frac12 \mbox{Tr}\left(
{\cal Q}_{12}{\cal Q}_{22}^{-1}{\cal Q}_{21}{\cal Q}_{11}^{-1} \right).
\label{ApprL}
\ee
But each matrix ${\cal Q}_{kk}^{-1}$ ($k=1,3$) contains the denominator
$\det{\cal Q}_{kk}$, which can be much less than any element of matrix
${\cal Q}_{kk}$. If this happens, then the inequality
$\mbox{Tr}\left(
{\cal Q}_{12}{\cal Q}_{22}^{-1}{\cal Q}_{21}{\cal Q}_{11}^{-1}\right)
\gg \mbox{Tr}\left({\cal Q}_{12}{\cal Q}_{21}\right)/
\left(\mbox{Tr}{\cal Q}_{11}\mbox{Tr}{\cal Q}_{22}\right)$
becomes quite possible.
Just such a situation takes place in the example considered.
Although diagonal elements $\overline{x_kx_k}$ and $\overline{p_kp_k}$ of
matrices ${\cal Q}_{kk}$ grow exponentially with time,
these matrices have
also exponentially growing off-diagonal covariance elements
$\widetilde{x_kp_k}$
(this means that each mode occurs in
{\em highly-correlated quantum state\/}
\cite{DKM80} with quadrature correlation coefficient
$r\equiv \widetilde{x_kp_k}/
\left(\overline{x_kx_k}\;\overline{p_kp_k}\right)^{1/2}$
approaching the unit value), so that $\det{\cal Q}_{kk}$ does not grow
unlimitedly with time, exhibiting only relatively small oscillations.
For this reason, elements of matrices ${\cal Q}_{kk}^{-1}$ have the
same order of magnitude ($\sim \exp(2\tau)\,$) as elements
of matrices ${\cal Q}_{kk}$ themselves. On the other hand, elements
of matrix ${\cal Q}_{12}$ have an order of $\exp(-2\tau)$ at the
moments of intermediate minima. Therefore, the exponential
time dependences are
canceled in the measures based on determinants, resulting in the
inequalities ${\cal J}_c, \widetilde{\cal L} \gg {\cal Y}_*$ for
$\cos(\rho\tau)\approx 0$.

This example permits us to make a conjecture that the
covariance entanglement coefficient ${\cal Y}$ is not only simpler
from the point of view of calculations, but it could be preferable from
the physical point of view, because it
is more sensitive to entanglement than entropic and purity measures.
Other arguments in favour of ${\cal Y}$ can be found in \cite{PLA}.

\subsection{Asymmetric resonance}

An interesting feature of the Hamiltonian (\ref{H2}) discovered in
\cite{Sasha} is a possibility to compensate one detuning (e.g.,
$\delta$) at the expense of another. In particular, an exponential
growth of the energies of both modes can be obtained under the
conditions of ``asymmetric resonance''
\be
\delta=\ep, \qquad
3\delta - \Delta  =\ep \nu/2 \,.
\label{asymm}
\ee
In this case the quadrature components depend on time as follows:
\beqn
x_1(t) &=& x_1(0)
\left[\left(1-\frac{2}{\nu}\right) C_{1}^{-}(2R\tau;t)
+\frac{2}{\nu}\cos\phi_1 \right]
- p_1(0)
\left[\left(1-\frac{2}{\nu}\right) S_{1}^{-}(2R\tau;t)
-\frac{2}{\nu}\sin\phi_1 \right]
\nonumber\\&&
+\frac{x_3(0)}{4\mu}
\left[C_{1}^{-}(2R\tau;t) -\cos\phi_1 \right]
-\frac{p_3(0)}{12\mu}
\left[S_{1}^{-}(2R\tau;t) +\sin\phi_1 \right] ,
\label{x1t1}
\eeqn
\beqn
x_3(t) &=& x_3(0)
\left[\left(1-\frac{2}{\nu}\right)\cos\phi_3
+\frac{2}{\nu} C_{3}^{-}(2R\tau;t) \right]
+\frac13 p_3(0)
\left[\left(1-\frac{2}{\nu}\right)\sin\phi_3
-\frac{2}{\nu} S_{3}^{-}(2R\tau;t) \right]
\nonumber\\&&
+\frac{x_1(0)}{12\mu}
\left[C_{3}^{-}(2R\tau;t) -\cos\phi_3 \right]
-\frac{p_1(0)}{12\mu}
\left[S_{3}^{-}(2R\tau;t) +\sin\phi_3 \right],
\label{x2t1}
\eeqn
\beqn
p_1(t) &=&- x_1(0)
\left[\left(1-\frac{2}{\nu}\right) S_{1}^{+}(2R\tau;t)
+\frac{2}{\nu}\sin\phi_1 \right]
+ p_1(0)
\left[\left(1-\frac{2}{\nu}\right) C_{1}^{+}(2R\tau;t)
+\frac{2}{\nu}\cos\phi_1 \right]
\nonumber\\&&
-\frac{x_3(0)}{4\mu}
\left[S_{1}^{+}(2R\tau;t) -\sin\phi_1 \right]
+\frac{p_3(0)}{12\mu}
\left[C_{1}^{+}(2R\tau;t) -\cos\phi_1 \right]  ,
\label{p1t1}
\eeqn
\beqn
p_3(t) &=& -3x_3(0)
\left[\left(1-\frac{2}{\nu}\right)\sin\phi_3
+\frac{2}{\nu} S_{3}^{+}(2R\tau;t) \right]
+ p_3(0)
\left[\left(1-\frac{2}{\nu}\right)\cos\phi_3
+\frac{2}{\nu} C_{3}^{+}(2R\tau;t) \right]
\nonumber\\&&
-\frac{x_1(0)}{4\mu}
\left[S_{3}^{+}(2R\tau;t) -\sin\phi_3 \right]
+\frac{p_1(0)}{4\mu}
\left[C_{3}^{+}(2R\tau;t) -\cos\phi_3 \right],
\label{p2t1}
\eeqn
where
\[
\phi_{k}(\tau;t)=k\overline{\omega} t - 2J\tau,
\qquad  R= 1-\frac{2}{\nu}\,, \qquad  J= \frac{\nu}{2} +1,
\]
and all terms of the order of ${\cal O}(\nu^{-2})$ have been neglected
(as well as the corrections of the order of $\delta\sim\ep$ in the amplitude
coefficients).

The (normalized) mean energies of each mode depend on time as
follows \cite{Sasha}:
\be
{\cal E}_1 = \frac{\theta_1}{2}\left[
\left(1-\frac{4}{\nu}\right)\cosh(4R\tau) + \frac{4}{\nu}\psi(\tau)
\right]
+ \frac{\theta_3}{\nu}\left[\cosh(4R\tau) + 1 -2\psi(\tau)\right],
\label{E1ofr}
\ee
\be
{\cal E}_3 = \frac{\theta_3}{2}\left[
1-\frac{4}{\nu} + \frac{4}{\nu}\psi(\tau)\right]
+ \frac{\theta_1}{\nu}\left[\cosh(4R\tau) + 1 -2\psi(\tau)\right],
\label{E3ofr}
\ee
where
\[
\psi(\tau) \equiv \cosh(2R\tau)\cos(2J\tau).
\]
The energy of the third mode is significantly less than the energy of the
first mode, if $\nu\gg 1$.
For this reason this regime of excitation was named ``asymmetrical''.
For $\tau>1$, ${\cal E}_3/ {\cal E}_1\approx 6/\nu$.
Note, however, that for the cubical
cavity with $\nu=50/3$, the energy of the third mode is only three times
less than that of the first one.
It is important, nonetheless, that
the rates of increase of the energies of each mode are
almost twice bigger than they were in the case of the strict resonance
discussed in the preceding subsection.

The covariance entanglement coefficients can be written again in the
form  (\ref{Y13}), but with ${\cal E}_{1,3}$ given by (\ref{E1ofr}) and
(\ref{E3ofr}). The function ${\cal F}$ in the asymmetric case reads
(neglecting corrections of the order of $\nu^{-2}$ with respect to the
main terms)
\beqn
{\cal F} &=& 2\nu^{-1}\Bigg\{
\theta_1^2\Big( \cosh^2(4R\tau)
+ \sinh^2(2R\tau) - \cosh(6R\tau)\cos\phi_0
 \nonumber \\ &&
+ 2\nu^{-1}\left[\cos\phi_0 \left\{ \cosh(2R\tau) +3\cosh(6R\tau)\right\}
-2\cosh^2(2R\tau)\cos^2\phi_0 -2\cosh(4R\tau) -2\sinh^2(4R\tau)\right]
\Big)
 \nonumber \\ &&
+\theta_3^2\Big( \cosh^2(2R\tau)
 - \cosh(2R\tau)\cos\phi_0
 \nonumber \\ &&
+ 2\nu^{-1}\left[\cos\phi_0 \left\{ \cosh(6R\tau) +3\cosh(2R\tau)\right\}
-2\cosh^2(2R\tau)\cos^2\phi_0 -2\cosh(4R\tau) \right]
\Big)
 \nonumber \\ &&
+2\theta_1\theta_3\Big( \cosh(4R\tau)\left[
 \cosh(2R\tau)\cos\phi_0 -1\right]
 \nonumber \\ &&
+ 2\nu^{-1}\left[-2\cos\phi_0 \left\{ \cosh(6R\tau) +\cosh(2R\tau)\right\}
+2\cosh^2(2R\tau)\cos^2\phi_0 + 4\cosh^4(2R\tau) -2
\right]
\Big)
\Bigg\},
\label{F-asym}
\eeqn
where $\phi_0=-2J\tau$.

If $\tau \to \infty$, then (for $\nu \gg 1$)
\[
{\cal F} \approx \frac{\theta_1^2}{2\nu} e^{8R\tau}, \qquad
{\cal E}_1 \approx \frac{\theta_1}{4} e^{4R\tau}, \qquad
{\cal E}_3 \approx \frac{\theta_1}{2\nu} e^{4R\tau},
\]
so that ${\cal Y}\to 1$, whereas $\tilde{\cal Y}\to \sqrt{8/\nu}$.
Consequently, the coefficient ${\cal Y}$ is preferable when the
energies of subsystems are essentially different.

The purity and entropic entanglement coefficients are given by
Eqs. (\ref{Ltil13}) and (\ref{entr13}), with
\be
g_{1}^2(\tau) =
1 +\frac{8}{\nu}\left[\left(1-\theta_{31}\right)\psi(\tau) -1
+ \theta_{31}\cosh^2(2R\tau) \right]
\label{IUP1}
\ee
and $g_{3}$ obtained from (\ref{IUP1}) by means of the replacement
$1\leftrightarrow 3$.
We see a significant difference from the strict resonance case: now
functions $g_{1,3}(\tau)$ increase exponentially with time for $\tau\gg 1$.
Asymptotically, each mode appears in a highly mixed
quantum state, with
$\det{\cal Q}_{11}= \det{\cal Q}_{33}=\theta_1\theta_3\exp(4R\tau)/(2\nu)$.
The purity entanglement coefficient (\ref{Ltil13})
tends asymptotically to the unit value independently of the initial
temperature (or coefficients $\theta_k$):
\[
\widetilde{\cal L} \approx 1- \frac{\nu}{2}\exp(-4R\tau), \qquad \tau\gg 1.
\]
For $\tau\gg 1$ the entropic entanglement coefficient grows unlimitedly:
$I_c \sim \ln (g_1g_3) \sim 4R\tau$. Therefore in Fig.~\ref{figYJLasym}
we compare the compact parameter
${\cal J}_c(\tau)$ (\ref{defJcnew}) with the functions
${\cal Y}(\tau)$ (\ref{Y}),
$\widetilde{\cal L}$ (\ref{Ltil13}), and
$[\widetilde{\cal L}(\tau)]^{1/2} $ for the initial vacuum state.
Since all formulae in the asymmetric case are obtained
neglecting terms of the order of $\nu^{-2}$, we use the value $\nu=100$
in the illustrations. The difference between the asymptotical values
of the functions for $\tau\gg 1$ and the correct value $1$
shows the accuracy of approximation (about $2$\%).
The dependences of the entanglement covariance and purity
coefficients ${\cal Y}$ and $\widetilde{\cal L}$ on the ``slow time'' $\tau$
for the initial vacuum and high-temperature state
are shown in Fig. \ref{figY-Lasym}. Remember that in the high-temperature
case the coefficient $\widetilde{\cal L}$ tends to the compact entropic
coefficient ${\cal J}_c$.

\section{Fabry-Perot cavity with an oscillating boundary}

The problem of the scalar massless field in a 1D cavity formed by
two infinite ideal plates whose positions are given by $x_{left}\equiv 0$
and 
\begin{equation}
x_{right}\equiv L(t)=L_0\left(1+\varepsilon \sin\left[p\omega_1 t\right]
\right),
\quad |\varepsilon|\ll 1,
\quad  \omega_1=\pi c/L_0, \quad p=1,2,\ldots
\label{L(t)}
\end{equation}
was solved in \cite{Djpa}.
The only
component of the operator vector potential of the electromagnetic field $%
\hat{A}(x,t)$ \emph{in the Heisenberg representation\/}
can be written as
\begin{equation}
\hat{A}(x,t)=\sum_{n=1}^{\infty}\frac 2{\sqrt {n}}\left[ \hat
b_n\psi^{(n)}(x,t)\,+\,\mbox{h.c.}\,\right], \qquad \left[\hat b_n\,,\,\hat
b_k^{\dagger}\right]=\delta_{nk}\, ,  \label{vecpot}
\end{equation}
where
\begin{equation}
\psi^{(n)}(x,t)=\sqrt {\frac{L_0}{L(t)}}\sum_{k=1}^{\infty}
 \sin \left[\frac{\pi kx}{L(t)}\right]
\left\{
\rho_k^{(n)}(\tau)e^{-i\omega_k t}
-\rho_{-k}^{(n)}(\tau)e^{i\omega_k t}
\right\},
\label{decom}
\end{equation}
\be
\tau =\frac 12\varepsilon\omega_1t, \qquad
\omega_n =n\omega_1.
\label{deftau1d}
\ee
The normalization
factors $2/\sqrt{n}$ in (\ref{vecpot}) are chosen in such a way that the
energy of the field in the stationary case can be represented as a sum of
energies of independent mode oscillators.
The coefficients $\rho_k^{(n)}(\tau)$
satisfy an infinite system of coupled equations
($k=\pm1,\pm2,\ldots$; $n=1,2,\ldots $)
\begin{equation}
\frac {\mbox{d}}{\mbox{d}\tau}\rho_k^{(n)}= \sigma\left[(k+p)%
\rho_{k+p}^{(n)}- (k-p)\rho_{k-p}^{(n)}\right] \,,
\qquad \sigma\equiv (-1)^p,
\label{prhok}
\end{equation}
which was solved in \cite{Djpa}
(here we confine ourselves to the simplest special case of
solutions found in \cite{Djpa}, corresponding to the {\em strict\/}
resonance).

Due to the initial conditions $\rho_k^{(n)}(0)=\delta_{kn}$ the solutions to
(\ref{prhok}) satisfy the relation $\rho_{j+mp}^{(k+np)}\equiv 0$ if $j\neq
k $. The non-zero coefficients $\rho_m^{(n)}$ read \cite{Djpa} 
\be
\rho_{j+mp}^{(j+np)}(\tau)= \frac{\Gamma\left(1+n+j/p\right)
(\sigma\kappa)^{n-m}} {\Gamma\left(1+m+j/p\right)\Gamma%
\left(1+n-m\right) }
 F\left(n+j/p\,,-m -j/p\,; 1+n-m\,; \kappa^2\right),
 \label{rho2}
\ee
where
\be
\kappa=\tanh(p\tau)
\label{def-kappa}
\ee
and $F(a,b;c;z)$ is the Gauss hypergeometric function.
The functions (\ref{rho2}) are \emph{exact\/} solutions to the set of
equations (\ref{prhok}) relating the coefficients with different lower
indices. Besides, these functions satisfy another set of equations, which
can be treated as recurrence relations with respect to the \textit{upper\/}
indices \cite{Djpa} 
\begin{equation}
\frac{d }{d\tau}\rho_m^{(n)} = n\left\{\sigma\left[ \rho_m^{(n-p)}
-\rho_m^{(n+p)}\right] \right\}, \quad n\ge p, \quad
\rho_m^{(0)}\equiv 0  \label{recrho}
\end{equation}
\begin{equation}
\frac{d }{d\tau}\rho_m^{(n)} = n\left\{\sigma\left[ \rho_{-m}^{(p-n)*}
-\rho_m^{(p+n)}\right] \right\}, \quad n=1,2,\ldots,p-1
\label{recrho1}
\end{equation}
The consequences of equations (\ref{prhok}), (\ref{recrho}) and (\ref%
{recrho1}) are the identities 
\begin{eqnarray}
&& \sum_{m=-\infty}^{\infty} m\rho_{m}^{(n)*}\rho_{m}^{(k)} =n\delta_{nk}\,,
\quad n,k=1,2,\ldots  \label{rhocond1} \\[2mm]
&& \sum_{n=1}^{\infty}\frac{m}{n} \left[\rho_{m}^{(n)*}\rho_{j}^{(n)} -
\rho_{-m}^{(n)*}\rho_{-j}^{(n)} \right] =\delta_{mj}\,, \quad m,j=1,2,\ldots
\label{rhocond2} \\[2mm]
&& \sum_{n=1}^{\infty}\frac{1}{n} \left[\rho_{m}^{(n)*}\rho_{-j}^{(n)} -
\rho_{j}^{(n)*}\rho_{-m}^{(n)} \right] =0\,, \quad m,j=1,2,\ldots
\label{rhocond3}
\end{eqnarray}

We suppose that after some interval of time $T$ the wall comes back to its
initial position $L_{0}$. For $t\geq T$, the field operator assumes the form 
\begin{equation}
\hat{A}(x,t)=\sum_{n=1}^{\infty }\frac{2}{\sqrt{n}}\sin \left( \pi
nx/L_{0}\right) \left[ \hat{a}_{n}e^{-i\omega _{n}t}\,+\,%
\mbox{h.c.}\,\right]  \label{vecpotfin}
\end{equation}%
where operators $\hat{a}_{m}$ are related to the initial operators $\hat{b}%
_{n}$ and $\hat{b}_{n}^{\dagger }$ by means of the Bogoliubov transformation
($\tau _{T}\equiv \frac{1}{2}\varepsilon \omega _{1}T$) 
\begin{equation}
\hat{a}_{m}=\sum_{n=1}^{\infty }\sqrt{\frac{m}{n}}\left[ \hat{b}_{n}\rho
_{m}^{(n)}\left( \tau _{T}\right) -\hat{b}_{n}^{\dag }\rho _{-m}^{(n)\ast
}\left( \tau _{T}\right) \right] ,\quad m=1,2,\ldots .  \label{Bogol}
\end{equation}%
The commutation relations $\left[ \hat{a}_{n}\,,\,\hat{a}_{k}^{\dagger }%
\right] =\delta _{nk}$ hold due to the identities (\ref{rhocond1})-(\ref%
{rhocond3}) which are nothing but the \textit{unitarity conditions\/} of the
transformation (\ref{Bogol}). These commutation relations together with the
expression for the energy of the field 
\begin{equation}
\hat{H}\equiv \frac{1}{8\pi }\int_{0}^{L_{0}}\mbox{d}x\,\left[ \left( _{{}}%
\frac{\partial \hat{A}}{\partial t}\right) ^{2}+\left( _{{}}\frac{\partial 
\hat{A}}{\partial x}\right) ^{2}\right] =\sum_{n=1}^{\infty }\omega
_{n}\left( \hat{a}_{n}^{\dag }\hat{a}_{n}+\frac{1}{2}\right)  \label{Ham}
\end{equation}%
convince us that $\hat{a}_{n}$ and $\hat{a}_{n}^{\dag }$ are true
photon annihilation and creation operators at $t\geq T$ (like the operators $%
\hat{b}_{n}$ and $\hat{b}_{n}^{\dag }$ were `physical' ones at $t<0$).

\subsection{Intermode entanglement in the parametric resonance case
($p=2$)}

Our first goal is to calculate the entanglement coefficients between
different modes in the case of the parametric resonance, $p=2$.
If the initial state of the field was vacuum
\textit{with respect to the initial operators\/} $\hat{b}_{n}$:
$\hat{b}_{n}|0\rangle =0$ (we use here the Heisenberg picture),
then the covariance entanglement coefficient between
the $r$th and $s$th modes is
\begin{equation}
{\cal Y}_{r,s}=\left[ \frac{|\langle \hat{a}_{r}\hat{a}_{s}\rangle|
^{2}+|\langle \hat{a}_{r}^{\dagger }\hat{a}_{s}\rangle| ^{2}}
{2\left(\langle \hat{a}_{r}^{\dagger }\hat{a}_{r}\rangle +1/2\right)
\left(\langle \hat{a}_{s}^{\dagger }\hat{a}_{s}\rangle +1/2\right) }
\right] ^{1/2}.
\label{Yrsdef}
\end{equation}
Using (\ref{Bogol}), one can express
the average values contained in formula (\ref{Yrsdef})
as (assuming hereafter $\omega _{1}=1$)
\begin{equation}
\langle \hat{a}_{r}\hat{a}_{s}\rangle
=-\sqrt{rs}\sum_{n=1}^{\infty }%
\frac1{n}{\rho _{r}^{(n)}\rho _{-s}^{(n)\ast }}
=-\sqrt{rs}\sum_{n=1}^{\infty }%
\frac1{n}{\rho _{s}^{(n)}\rho _{-r}^{(n)\ast }},
\label{var}
\end{equation}%
\be
 \langle \hat{a}_{r}^{\dagger }\hat{a}_{s}\rangle =
\sqrt{rs}\sum_{n=1}^{\infty }\frac1{n}{\rho
_{-r}^{(n)}\rho _{-s}^{(n)\ast }},
\qquad
\langle \hat{a}_{r}^{\dagger }\hat{a}_{r}\rangle
=r\sum_{n=1}^{\infty }\frac1{n}\left| \rho _{-r}^{(n)}\right| ^{2},
\label{covar}
\end{equation}%
where the coefficients $\rho _{\pm m}^{(n)}$ should be taken at the moment $%
T $, thus their argument is $\tau _{T}$. Strictly speaking, the expressions
in (\ref{var}) and (\ref{covar}) have physical meanings at those moments
of time $T$ when the wall returns to its initial position, i.e. for
$T=N\pi /p$ with an integer $N$. Consequently, the argument $\tau _{T}$
of the coefficients $\rho _{\pm m}^{(n)}$ in (\ref{var}) and (\ref{covar})
assumes discrete values $\tau ^{(N)}=N\varepsilon \pi /(2p)$. One should
remember, however, that something interesting in our problem happens for the
values $\tau \sim 1$ (or larger). Then $N\sim \varepsilon ^{-1}\gg 1$, and
the minimal increment $\Delta \tau \sim \varepsilon $ is so small that $\tau
_{T}$ can be considered as a continuous variable (under the realistic
conditions, $\varepsilon \leq 10^{-8}$ \cite{DK96}). For this reason, we
omit hereafter the subscript $T$, writing simply $\tau $ instead of $\tau
_{T}$ or $\tau ^{(N)}$.

Differentiating the right-hand sides of equations (\ref{var}) and
(\ref{covar})
with respect to the `slow time' $\tau $, one can remove the
fraction $1/n$ with the aid of the recurrence relations (\ref{recrho}) and (%
\ref{recrho1}). After that, changing if necessary the summation index $n$ to 
$n\pm p$, one can verify that almost all terms in the right-hand sides are
cancelled, and the infinite series are reduced to the finite sums.
For $p=2$ we obtain the equations (taking into account that all functions
$\rho_{m}^{(n)}$ are {\em real\/} in the strict resonance case, according
to Eq. (\ref{rho2})\,)
\begin{equation}
\mbox{d}\langle \hat{a}_{r}\hat{a}_{s}\rangle /\mbox{d}\tau =
- \sqrt{rs}\left[ \rho _{r}^{(1)}\rho _{s}^{(1)}+\rho
_{-r}^{(1) }\rho _{-s}^{(1) }\right] ,  \label{vmedio1}
\end{equation}%
\begin{equation}
\mbox{d}\langle \hat{a}_{r}^{\dagger }\hat{a}_{s}\rangle /\mbox{d}\tau
= \sqrt{rs}\left[ \rho _{r}^{(1) }\rho_{-s}^{(1) }
+\rho _{-r}^{(1)}\rho _{s}^{(1)}\right] ,
\qquad
\mbox{d}\langle \hat{a}_{r}^{\dagger }\hat{a}_{r}\rangle/\mbox{d}\tau
=2r\, \rho _{r}^{(1)}\rho _{-r}^{(1)} .
\label{vmedio3}
\end{equation}

For $p=2$, only odd modes can be excited from the initial vacuum state.
In this case, the hypergeometric functions in the formula (\ref{rho2})
for coefficients $\rho _{r}^{(n)}$
with $j=1$ are reduced to some combinations of the complete elliptic
integrals of the first and the second kinds \cite{Djpa}
\[
{\bf K}(\kappa )=\int_0^{\pi /2}\frac {\mbox{d}\alpha}{\sqrt {1
-\kappa^2\sin^2\alpha}}
=\frac{\pi}{2} F\left(\frac12\,,\,\frac12\,;\,1\,;\,\kappa^2\right),
\]
\[
{\bf E}(\kappa )=\int_0^{\pi /2}\mbox{d}\alpha
\sqrt {1-\kappa^2\sin^2\alpha}
=\frac{\pi}{2} F\left(-\frac12\,,\,\frac12\,;\,1\,;\,\kappa^2\right),
\]
so that equations (\ref{vmedio1}) and (\ref{vmedio3})
can be integrated for any values of $r$ and $s$:
see \cite{Djpa,DA} or Appendix \ref{ap-int} for technical details.
In particular, for the first few modes we find
\be
\langle \hat{a}_{1}^{2}\,\rangle =\frac{2}{\pi ^{2}\kappa }\left[ \tilde{%
\kappa}^{2}\mathbf{K}^{2}-2\mathbf{EK}+\mathbf{E}^{2}\right],
\label{a1quad}
\ee
\be
\langle \hat{a}_{3}^{2}\,\,\rangle =\frac{2}{9\pi ^{2}\kappa ^{3}}\left[ 
\tilde{\kappa}^{2}(4-\kappa ^{2})\mathbf{K}^{2}-2(2\kappa ^{4}-3\kappa
^{2}+4)\mathbf{EK}+(4\kappa ^{4}-\kappa ^{2}+4)\mathbf{E}^{2}\right],
\label{a3quad}
\ee
\begin{equation}
\langle \hat{a}_{1}\,\hat{a}_{3}\,\rangle =-\frac{2\sqrt{3}}{3\pi ^{2}\kappa
^{2}}\left[ \tilde{\kappa}^{2}\mathbf{K}^{2}-2\mathbf{EK}+\left( 1+\kappa
^{2}\right) \mathbf{E}^{2}\right],  \label{mcross1}
\end{equation}%
\begin{equation}
\langle \hat{a}_{1}^{\dagger }\hat{a}_{3}\,\rangle =\frac{2\sqrt{3}}{\pi
^{2}\kappa }\left[ \frac{\tilde{\kappa}^{2}}{3}\mathbf{K}^{2}+\frac{2}{3}%
(\kappa ^{2}-2)\mathbf{EK}+\mathbf{E}^{2}\right],  \label{mcross2}
\end{equation}%
\begin{equation}
\langle \hat{a}_{1}\,\hat{a}_{5}\,\rangle =\frac{2\sqrt{5}}{45\pi ^{2}\kappa
^{3}}\left[ \tilde{\kappa}^{2}(\kappa ^{2}+8)\mathbf{K}^{2}-2(\kappa ^{4}+8)%
\mathbf{EK}+(8\kappa ^{4}+7\kappa ^{2}+8)\mathbf{E}^{2}\right],
\label{mcross3}
\end{equation}%
\begin{equation}
\langle \hat{a}_{1}^{\dagger }\hat{a}_{5}\,\rangle =-\frac{2\sqrt{5}}{3\pi
^{2}\kappa ^{2}}\left[ \frac{\tilde{\kappa}^{2}}{5}(2\kappa ^{2}+1)\mathbf{K}%
^{2}+\frac{2}{5}(2\kappa ^{4}-2\kappa ^{2}-3)\mathbf{EK}+(\kappa ^{2}+1)%
\mathbf{E}^{2}\right],  \label{mcross4}
\end{equation}%
\begin{eqnarray}
\langle \hat{a}_{3}\,\hat{a}_{5}\,\rangle &=&\frac{2\sqrt{15}}{45\pi
^{2}\kappa ^{4}}\left[ \tilde{\kappa}^{2}(\kappa ^{2}+2)(\kappa ^{2}-2)%
\mathbf{K}^{2}+2(2\kappa ^{6}-\kappa ^{4}-2\kappa ^{2}+4)\mathbf{EK}\right. 
\nonumber \\
&&\left. -4(\kappa ^{2}+1)(\kappa ^{4}-\kappa ^{2}+1)\mathbf{E}^{2}\right],
\label{mcross5}
\end{eqnarray}%
\begin{equation}
\langle \hat{a}_{3}^{\dagger }\hat{a}_{5}\,\rangle =\frac{2\sqrt{15}}{45\pi
^{2}\kappa ^{3}}\left[ \tilde{\kappa}^{2}(7\kappa ^{2}-4)\mathbf{K}%
^{2}+2(8\kappa ^{4}-15\kappa ^{2}+4)\mathbf{EK}-(4\kappa ^{4}-19\kappa
^{2}+4)\mathbf{E}^{2}\right],  \label{mcross6}
\end{equation}%
\begin{equation}
\mathcal{E}_{1}=\frac{2}{\pi ^{2}}\mathbf{K}\left( 2\mathbf{E}-\tilde{\kappa}%
^{2}\mathbf{K}\right),  \label{mcross7}
\end{equation}%
\begin{equation}
{\cal E}_{3}=\frac{2}{3\pi ^{2}\kappa ^{2}}\left[ \left( 3\kappa
^{2}-2\right) \mathbf{K}\left( 2\mathbf{E}-\tilde{\kappa}^{2}\mathbf{K}%
\right) +2\left( 1+\kappa ^{2}\right) \mathbf{E}^{2}\right] \,
\label{mcross8}
\end{equation}%
\begin{eqnarray}
{\cal E}_{5} &=&-\frac{2}{45\pi ^{2}\kappa ^{4}}\left[ \tilde{\kappa}%
^{2}(47\kappa ^{4}-30\kappa ^{2}-8)\mathbf{K}^{2}+2(4\kappa ^{6}-47\kappa
^{4}+26\kappa ^{2}+8)\mathbf{EK}\right. \,  \nonumber \\
&&\left. -2(\kappa ^{2}+1)(4\kappa ^{4}+11\kappa ^{2}+4)
\mathbf{E}^{2}\right],
\label{mcross9}
\end{eqnarray}%
where $\tilde{\kappa}\equiv \sqrt{1-\kappa ^{2}}$ and we used
${\cal E}_{r}=\langle \hat{a}_{r}^{\dagger }\hat{a}_{r}\rangle +1/2$.

In Figure \ref{figY13}
we show ${\cal Y}_{1,3}$ and ${\cal Y}_{3,5}.$ We see
that the entanglement is strongest for the lowest modes.
However, for any pair $r,s$ the coefficient ${\cal Y}_{r,s}$ tends
asymptotically to the {\em unit\/} value when $\kappa\to 1$.
To prove this property, one should use the asymptotical
forms of the coefficients $\rho _{m}^{(n)}$
for $\tau\to \infty$, i.e., for $\kappa\to 1$.
Namely, replacing the hypergeometric functions in (\ref{rho2}) by
their values for the unit argument \cite{Grad},
\[
F(a,b;a+b+1;1)=\frac{\Gamma (a+b+1)}{\Gamma (a+1)\Gamma (b+1)}\,,
\]
one obtains the following asymptotical formulae
(see also \cite{DA}):
\be
\rho _{2m+1}^{(1)}(\tau) = \rho _{-2m-1}^{(1)}(\tau)=
\frac{2(-1)^m}{\pi(2m+1)}, \qquad \tau \to \infty.
\label{asrho}
\ee
Consequently, for $\tau \gg 1$ we have
\be
\langle \hat{a}_{r}^{\dagger }\hat{a}_{s}\rangle \approx
-\langle \hat{a}_{r}\hat{a}_{s}\rangle \approx
\frac{8\tau}{\pi^2\sqrt{rs}}(-1)^{(r-s)/2} + {\cal O}(1),
\label{asymp}
\ee
and the leading terms in the numerator and denominator of the fraction
in (\ref{Yrsdef}) become the same for $\tau\to\infty$.

In the case of detuning from the strict resonance, characterized by some
dimensionless detuning parameter $\gamma$,
the coefficients $\rho _{m}^{(n)}$
become complex. However, their asymptotical forms differ from
(\ref{asrho}) only by some phase factors \cite{DA}.
Since the covariance entanglement coefficient (\ref{Yrsdef}) depends on the
{\em absolute values\/} of the second-order moments
$\langle \hat{a}_{r}^{\dagger }\hat{a}_{s}\rangle$ and
$\langle \hat{a}_{r}\hat{a}_{s}\rangle$, these phase
factors do not influence the final result, namely,
that ${\cal Y}_{rs}\to 1$ when $\tau\to\infty$,
unless the dimensionless detuning parameter exceeds the critical
value $\gamma=1$,
when the generation of photons from vacuum becomes impossible.

In the case of the initial vacuum state, the state of the field at the
subsequent moments of time remains {\em Gaussian\/} \cite{DA}, and
the purity entanglement coefficient can be calculated by means of
formula (\ref{purenGaus}). In a generic case, the determinant
of the symmetrical $4\times4$ matrix ${\cal Q}$ (\ref{univD0})
contains $17$ different
terms. However, in the specific case involved all
covariances between the ``coordinate'' and ``momenta'' operators turn
out to be equal to zero identically:
$\widetilde{x_kp_j}  =0$,
and for this reason the determinant
of the covariance matrix for the $i$th and $j$th modes
can be factorized in the following simple form:
\be
\det Q=\left(\sigma _{p_{i}p_{i}}\sigma_{p_{j}p_{j}}
-\sigma _{p_{i}p_{j}}^{2}\right)
\left(\sigma _{x_{i}x_{i}}\sigma_{x_{j}x_{j}}
-\sigma _{x_{i}x_{j}}^{2}\right).
\label{detp2}
\ee
Nonzero covariances are given by the following expressions:
\be
\sigma _{x_{i}x_{j}} \equiv \overline{x_{i}x_{j}}
= \frac12\langle
\hat{a}_{i}^{\dagger }\hat{a}_{j} + \hat{a}_{j}^{\dagger }\hat{a}_{i}
\rangle
+\mbox{Re}\langle \hat{a}_{i}\hat{a}_{j}\rangle ,
\qquad
\sigma _{p_{i}p_{j}} \equiv \overline{p_{i}p_{j}}
= \frac12\langle
\hat{a}_{i}^{\dagger }\hat{a}_{j} + \hat{a}_{j}^{\dagger }\hat{a}_{i}
\rangle
-\mbox{Re}\langle \hat{a}_{i}\hat{a}_{j}\rangle .
\label{varnonzer}
\ee
Introducing the correlation coefficients,
\be
r_{x_{i}x_{j}}=\frac{\sigma _{x_{i}x_{j}}}
{\sqrt{\sigma _{x_{i}x_{i}}\sigma _{x_{j}x_{j}}}}\,, \qquad
r_{p_{i}p_{j}}=\frac{\sigma _{p_{i}p_{j}}}
{\sqrt{\sigma _{p_{i}p_{i}}\sigma _{p_{j}p_{j}}}}\,,
\label{corcoef}
\ee
we can represent the $\widetilde{\cal L}$ (\ref{purenGaus})
and ${\cal Z}$ (\ref{ZGaus}) entanglement
coefficients between the $i$th and $j$th modes as
\be
\widetilde{\cal L}_{ij}=1 -
\sqrt{\left(1-r^2_{x_{i}x_{j}}\right)\left(1-r^2_{p_{i}p_{j}}\right)}\,,
\label{Lp2}
\ee
\be
{\cal Z}_{ij}=1 +
\sqrt{\left(1-r^2_{x_{i}x_{j}}\right)\left(1-r^2_{p_{i}p_{j}}\right)}
-2\,\sqrt{\frac
{\left(1-r^2_{x_{i}x_{j}}\right)\left(1-r^2_{p_{i}p_{j}}\right)}
{\left(1-\frac14 r^2_{x_{i}x_{j}}\right)\left(1-\frac14 r^2_{p_{i}p_{j}}
\right)} }\,.
\label{Zp2}
\ee
If all correlation coefficients are small (in particular, if $\tau \ll 1$),
then
\[
\widetilde{\cal L}_{ij} \approx 2{\cal Z}_{ij} \approx
\frac12 \left(r^2_{x_{i}x_{j}} + r^2_{p_{i}p_{j}}\right).
\]
When $\tau\to\infty$, then, due to equations (\ref{asymp}), (\ref{varnonzer})
and (\ref{corcoef}), the coefficients $\sigma _{p_{i}p_{j}}$ linearly
grow with time in such a way that the momentum correlation coefficient
$r_{p_{i}p_{j}}$ tends to the unit value.
At the same time, the coefficients $\sigma _{x_{i}x_{j}}$ and
$r_{x_{i}x_{j}}$ tend to some finite limit values.
Therefore, the purity entanglement coefficient $\widetilde{\cal L}$
and the distance entanglement coefficient ${\cal Z}$ approach the unit
value.
Using the asymptotical formulae for the complete elliptic
integrals \cite{Grad},
\[
\mathbf{K}(\kappa ) \approx \ln \frac{4}{\tilde{\kappa}}+\frac{1}{4}\left(
\ln \frac{4}{\tilde{\kappa}}-1\right) \tilde{\kappa}^{2}+\cdots\,,
\qquad
\mathbf{E}(\kappa ) \approx 1+\frac{1}{2}\left( \ln \frac{4}{\tilde{\kappa}%
}-\frac{1}{2}\right) \tilde{\kappa}^{2}+\cdots \,,
\qquad \kappa \to 1,
\]
one can see that for $\tau \gg 1$,
$1-\widetilde{\cal L} \sim 1- {\cal Z} \sim \tau^{-1/2}$.
In particular,
\[
1-\widetilde{\cal L}_{13}
\sim\sqrt{\frac{44}{57\tau}} \approx \frac{0.88}{\sqrt\tau}\,, \qquad
1- {\cal Z}_{13}
\sim \sqrt{\frac{44}{3\tau}}
\left(\frac{8}{\sqrt{219}} - \frac{1}{\sqrt{19}}\right)
\approx \frac{1.19}{\sqrt\tau}\,.
\]

Calculating the entropic entanglement measure (\ref{defIc}) one should
take into account that the reduced entropy of any two-mode subsystem depends
on time in the case involved (in contradistinction to the case considered
in the preceding section), because the evolution of each finite-dimensional
subsystem is not unitary. This entropy is determined by two eigenvalues
of the corresponding $4\times4$ matrix ${\cal Q}\,\Omega^{-1}$,
which are given by formula (\ref{solchareq}). The reduced
entropy of the $k$th mode is determined by the single number
\be
f_k=\sqrt{\sigma _{p_{k}p_{k}}\sigma _{x_{k}x_{k}}}\,,
\label{fk}
\ee
as soon as the
coordinate-momentum covariances are equal to zero in the strict resonance
case considered. The explicit formula for the entropic entanglement measure
between the $k$th and $n$th modes becomes (for the initial vacuum state
of the field)
\beqn
I_c^{kn} &=&
\sum_{j=k,n}\Big[ \left(f_j+1/2\right)\ln
\left(f_j+1/2\right) -
\left(f_j-1/2\right)\ln\left(f_j-1/2\right)\Big]
\nonumber \\ &&
 -\sum_{\delta=\pm 1}\Big[\left(f_{kn}^{\delta}+1/2\right)\ln
\left(f_{kn}^{\delta}+1/2\right) -
\left(f_{kn}^{\delta}-1/2\right)\ln\left(f_{kn}^{\delta}-1/2\right)\Big],
\label{Skn}
\eeqn
where
\beqn
\!\!\!\!\!\!\!\!\!\!
&&2f_{kn}^{\delta} =
\left[\overline{p_{k}p_{k}}\;\overline{x_{k}x_{k}}
+\overline{p_{n}p_{n}}\;\overline{x_{n}x_{n}}
+2\overline{p_{k}p_{n}}\;\overline{x_{k}x_{n}}
+2\,\sqrt{
\left(\overline{p_{k}p_{k}}\;\overline{p_{n}p_{n}}
-\overline{p_{k}p_{n}}^{2}\right)
\left(\overline{x_{k}x_{k}}\;\overline{x_{n}x_{n}}
-\overline{x_{k}x_{n}}^{2}\right) }\right]^{1/2}
\nonumber \\
\!\!\!\!\!
&&+
\delta \left[\overline{p_{k}p_{k}}\;\overline{x_{k}x_{k}}
+\overline{p_{n}p_{n}}\;\overline{x_{n}x_{n}}
+2\overline{p_{k}p_{n}}\;\overline{x_{k}x_{n}}
-2\,\sqrt{
\left(\overline{p_{k}p_{k}}\;\overline{p_{n}p_{n}}
-\overline{p_{k}p_{n}}^{2}\right)
\left(\overline{x_{k}x_{k}}\;\overline{x_{n}x_{n}}
-\overline{x_{k}x_{n}}^{2}\right) }\right]^{1/2}.
\label{fkn}
\eeqn
The behaviour of different entanglement coefficients is shown in
Fig. \ref{figYLJZ2p}. All of them monotonously tend to unity with
the course of time, but much more slowly than in the case of
asymmetric resonance in the 3D cavity (due to interaction with other
resonant modes).

\subsection{Entanglement in the ``semi-resonance case'' ($p=1$)}

A qualitatively different behaviour of all characteristics of the field
is observed in the  ``semi-resonance case'', when the frequency of the
oscillations of the boundary {\em coincides\/} with the fundamental field
eigenfrequency ($p=1$) \cite{Djpa,D96}.
In this case one should put $j=0$ in formula (\ref{rho2}),
and all coefficients $\rho _{m}^{(n)}$ with {\em negative\/}
lower indices $m$ are equal to zero identically. As a consequence,
no photons can be created from the initial vacuum state, which is
clearly seen from equation (\ref{covar}).
If initially the field was in non-vacuum state (at least for some mode),
then the total number of photons in all modes is conserved,
although the total energy grows exponentially due to ``heating''
the high-frequency modes (at the expense of ``cooling''
the low-frequency modes).

We suppose that initially only the first mode was excited, while all the
others were in the vacuum state. Then the dynamics of all modes is
described by means of the unique coefficient
\[
\rho _{m}^{(1)}={(\tanh \tau )^{m-1}}/{\cosh ^{2}\tau }.
\]
If initially the excited mode was in a coherent state, then
all second-order central moments connecting different modes
are equal to zero, resulting in zero covariance entanglement coefficient:
${\cal Y}_{r,s}^{coh}=0$.
However, for other initial states we obtain nonzero values of
${\cal Y}_{r,s}$.

If initially the first mode was in the Fock state $|n\rangle$, then
\be
{\cal Y}_{r,s}^{Fock}= \frac{n\zeta _{r}\zeta _{s}}
{\sqrt{2\left(n\zeta _{r}^{2}+1/2\right)
\left(n\zeta _{s}^{2}+1/2\right)}}\,,
\label{YFock}
\ee
where
\be
\zeta_m =\sqrt{m}\,\rho _{m}^{(1)}= \sqrt{m}\,
\frac{(\tanh \tau )^{m-1}}{\cosh ^{2}\tau } \le 1.
\label{rhom1}
\ee

If initially the first mode was in a squeezed vacuum state
$|\psi\rangle=
\exp \left[R(\hat{b}_{1}^{\dag 2}-\hat{b}_{1}^{2})/2\right]|0\rangle$
with the average number of photons
$\nu_{1}=\sinh ^{2}(R)$, then
\be
{\cal Y}_{r,s}^{sqz}=\frac{\zeta _{r}\zeta _{s}\sqrt{
\nu_{1}(2\nu_{1}+1)}}{\sqrt{2\left(\nu_{1}\zeta_{r}^{2}+1/2\right)
\left(\nu_{1}\zeta _{s}^{2}+1/2\right)}}\,,
\label{Ysqz1}
\ee

If initially the first mode was in an even/odd coherent state \cite{DMM74}
\[
|\alpha _{\pm}>=\frac{|\alpha _{1}\rangle \pm |-\alpha _{1}\rangle}
{\sqrt{2[1 \pm \exp (-2|\alpha_{1}|^{2})]}},
\]
then the mean numbers of photons are given by the formulae
\[
\nu_{1}^{(+)}=|\alpha _{1}|^{2}\tanh (|\alpha _{1}|^{2}), \qquad
\nu_{1}^{(-)}=|\alpha _{1}|^{2}\coth (|\alpha _{1}|^{2}).
\]
In both cases, the entanglement coefficient can be written as
\be
{\cal Y}_{r,s}^{ev/od}= \frac{\zeta _{r}\zeta _{s}\sqrt{
\nu_{1}\left(\nu_{1}+|\alpha _{1}|^{2}\right)}}
{\sqrt{2\left(\nu_{1}\zeta_{r}^{2}+1/2\right)
\left(\nu_{1}\zeta _{s}^{2}+1/2\right)}}\,.
\label{Yevod}
\ee
For big enough number of photons in the initial squeezed and even/odd states,
$\nu_1\gg 1$,
the entanglement coefficient becomes very close to the maximal possible
unit value, if $\nu_{1}\zeta_{r,s}^{2}\gg 1$, but with increase of time
${\cal Y}$ eventually goes to zero, because $\zeta_{r,s}(\tau)\to 0$
for $\tau\to\infty$. In the case of the initial Fock state, the maximal
value of ${\cal Y}$ does not exceed $1/\sqrt2$.
A typical behaviour of the covariance entanglement coefficient between the
first and second modes for the
initial Fock and squeezed states is shown in Fig. \ref{figY12}.
The behaviour of ${\cal Y}_{m,n}$ for the initial thermal and even/odd
states is very similar, especially for large mean numbers of photons.
The evolution of the mean number of photons in the first and second modes
is shown in Fig. \ref{figmeannum}.

The momentum-coordinate covariances turn out to be equal to zero again
(as in the case of $p=2$), therefore we need only two correlation
coefficients defined in (\ref{corcoef}), in order to calculate the
purity and distance entanglement coefficients (in the case of the initial
{\em squeezed\/} state of the first mode)
with the aid of Eqs.
(\ref{Lp2}) and (\ref{Zp2}). These correlation coefficients are as follows,
\be
r_{x_{i}x_{j}}=\frac{\chi \zeta_{i}(\tau)\zeta_{j}(\tau)}
{\sqrt{\left[1+\chi \zeta^2_{i}(\tau)\right]
\left[1+\chi \zeta^2_{j}(\tau)\right]}}\,, \qquad
r_{p_{i}p_{j}}=-\,\frac{\lambda \zeta_{i}(\tau)\zeta_{j}(\tau)}
{\sqrt{\left[1-\lambda \zeta^2_{i}(\tau)\right]
\left[1-\lambda \zeta^2_{j}(\tau)\right]}}\,,
\label{corqp1}
\ee
where
\[
\chi = e^{2R} -1, \qquad
\lambda = 1- e^{-2R}.
\]
The time dependences of the $\widetilde{\cal L}$ and ${\cal Z}$
entanglement coefficients are compared in Fig. \ref{figLZ1}.
We see that the full and dashed curves are very close,
especially for large mean numbers of photons.

\section{Conclusion}

We have compared time dependences of several functions characterizing
the degree of entanglement
between field modes of ideal cavities with resonantly vibrating walls
for different models of such cavities.
All these functions (the ``standard'' entropic entanglement
measure for Gaussian states, covariance entanglement coefficient
introduced in \cite{JRLR,PLA}, distance entanglement coefficient
introduced in \cite{MMSZ},
and purity entanglement coefficient introduced here)
are based on the second-order covariance matrix of the field
quadrature components. In spite of having different analytical forms,
the coefficients concerned show similar qualitative (and in certain cases
even quantitative) behaviour for each fixed model.
Therefore, the covariance entanglement coefficient,
being the simplest from the point of view of calculations,
seems to be the most convenient, especially compared with the entropic
entanglement measure, whose calculation requires tremendous efforts, giving
practically the same information on the degree of entanglement.
Moreover, an example at the end of section \ref{symres} shows that
the covariance entanglement coefficient
(based on {\em traces\/} of covariance submatrices)
can be more sensitive to entanglement than other measures
(which are based on {\em determinants\/} of covariance submatrices).

On the other hand, the behaviour of each selected
entanglement coefficient turns out
to be completely different for different kinds of cavities.
For the three-dimensional cavities with accidental degeneracy of the
spectrum of eigenfrequencies, the entanglement coefficients exhibit
oscillations in the case of ``symmetric'' resonance,
remaining relatively small for all instants of time.
Moreover, they go to zero periodically, despite that the energy of each mode
increases unlimitedly. In the case of ``asymmetric'' resonance, fast
(in the ``slow time'' scale) oscillations of the entanglement coefficients
are also observed, but all these coefficients tend to the maximal possible
unit value with increase of time.
For the model of one-dimensional (Fabry--Perot) cavity with equidistant
spectrum, all entanglement coefficients monotonously go to the unit value
in the parametric resonance case. In the ``semiresonance'' case, they
rapidly reach the values close to unity and remain at this level for
some interval of time (which increases with increase of the initial mean
number of quanta), but eventually they decay to zero.
Therefore, this study adds some new features to our understanding of the
behaviour of fields in cavities with vibrating boundaries, in addition
to results obtained earlier in \cite{DK96,Sasha,Djpa,DA,AD00}.

\section*{Acknowledgement}
The authors acknowledge a full support of the Brazilian agency CNPq.

\appendix
\section{The Bogoliubov coefficients in the 1D parametric resonance case}

\renewcommand{\theequation}{A.\arabic{equation}}
\setcounter{equation}{0}

The nonzero coefficients $\rho _{2m+1}^{(1)}$
in the parametric resonance case ($p=2$)
read \cite{Djpa,DA}
\begin{eqnarray}
\rho _{2m+1}^{(1)} &=&
\frac{(-1)^{m}\Gamma \left( m+1/2\right)
\kappa ^{m}}{\Gamma \left( 1/2\right) \Gamma \left(
1+m\right) }
F\left( m+1/2\,,\,-1/2\,;\,1+m\,;\,\kappa ^{2}\right) ,
 \label{ximn}
 \\
\rho _{-2m-1}^{(1)} &=&\frac{(-1)^{m}\Gamma \left( m+1/2\right)
\Gamma \left( 3/2\right) \kappa ^{m+1}}{\pi \Gamma \left(2+m\right) }
F\left( m+1/2\,,\,1/2\,\,;\,2+m\,;\,\kappa ^{2}\right) .
\label{etank}
\end{eqnarray}%
In particular
($\tilde{\kappa}\equiv \sqrt{1-\kappa ^{2}}$),
\begin{equation}
\rho _{1}^{(1)}=\frac{2}{\pi }\mathbf{E}(\kappa ),\qquad \rho _{-1}^{(1)}=%
\frac{2}{\pi \kappa }\left[ \mathbf{E}(\kappa )-\tilde{\kappa}^{2}\mathbf{K}%
(\kappa )\right],
\label{xiet1}
\end{equation}%
\begin{equation}
\rho _{3}^{(1)}=\frac{2}{3\pi \kappa }\left[ \left( 1-2\kappa ^{2}\right) 
\mathbf{E}(\kappa )-\tilde{\kappa}^{2}\mathbf{K}(\kappa )\right]
\qquad
\rho _{-3}^{(1)}=-\,\frac{2}{3\pi \kappa ^{2}}\left[ \left( 2-\kappa^{2}
\right) \mathbf{E}(\kappa )-2\tilde{\kappa}^{2}\mathbf{K}(\kappa )\right],
\label{et3}
\end{equation}%
\[
\rho _{5}^{(1)}=\frac{2}{15\pi \kappa ^{2}}\left[ \left( 8\kappa
^{4}-3\kappa ^{2}-2\right) \mathbf{E}(\kappa )+(-4\kappa ^{4}+2\kappa ^{2}+2)%
\mathbf{K}(\kappa )\right],
\]
\[
\rho _{-5}^{(1)}=-\frac{2}{15\pi \kappa ^{3}}\left[ \left( 2\kappa
^{4}+3\kappa ^{2}-8\right) \mathbf{E}(\kappa )-(\kappa ^{4}+7\kappa ^{2}-8)%
\mathbf{K}(\kappa )\right] .
\]
The general structure of the coefficients $\rho _{2m+1 }^{(1)}$
in terms of the complete elliptic integrals is \cite{DA}
\begin{equation}
\rho _{2m+1}^{(1)}=\frac{2}{\pi \kappa ^{m}}\left[
f_{m}\left( \kappa ^{2}\right) \mathbf{E}(\kappa )+\tilde{\kappa}%
^{2}g_{m}\left( \kappa ^{2}\right) \mathbf{K}(\kappa )\right]  \label{xim}
\end{equation}%
\begin{equation}
\rho _{-2m-1}^{(1)}=\frac{2}{\pi \kappa ^{m+1}}\left[
r_{m}\left( \kappa ^{2}\right) \mathbf{E}(\kappa )+\tilde{\kappa}%
^{2}s_{m}\left( \kappa ^{2}\right) \mathbf{K}(\kappa )\right]  \label{etm}
\end{equation}%
where $f_{m}(x),g_{m}(x),r_{m}(x),s_{m}(x)$ are polynomials of the
degree $m$ which can be found from the recurrence relations (\ref{prhok}).

\section{Calculation of integrals}

\label{ap-int}

To calculate, for instance, the average value
$\langle \hat{a}_{1}^{\dagger}\hat{a}_{3}\rangle $,
we use equations (\ref{vmedio3}), (\ref{xiet1}) and (\ref{et3}),
replacing the derivative over $\tau $ by the derivative with respect to $%
\kappa $ in accordance with the relation
$\mbox{d}\kappa =2\tilde{\kappa}^{2}\mbox{d}\tau $.
In this way we arrive at the equation
\be
\frac{\mbox{d}\langle \hat{a}_{1}^{\dagger }\hat{a}_{3}\,\rangle }{\mbox{d}%
\kappa } =-\frac{2\sqrt{3}}{3\pi ^{2}\kappa ^{2}\tilde{\kappa}^{2}}\left[
\left( 1+\kappa ^{2}\right) \mathbf{E}^{2}(\kappa )-\tilde{\kappa}^{4}%
\mathbf{K}^{2}(\kappa )
-2\kappa ^{2}\tilde{\kappa}^{2}\mathbf{E}(\kappa )\mathbf{K}(\kappa
)\right] .  \label{eqU1}
\ee
Taking into account the differentiation rules \cite{Grad} 
\begin{equation}
\frac{\mbox{d}\mathbf{K}(\kappa )}{\mbox{d}\kappa }=\frac{\mathbf{E}(\kappa )%
}{\kappa \tilde{\kappa}^{2}}-\frac{\mathbf{K}(\kappa )}{\kappa },
\qquad \frac{%
\mbox{d}\mathbf{E}(\kappa )}{\mbox{d}\kappa }=\frac{\mathbf{E}(\kappa )-%
\mathbf{K}(\kappa )}{\kappa },  \label{difrul1}
\end{equation}%
we may suppose that the factor $\tilde{\kappa}^{2}$ in the denominator of
the right-hand side of equation (\ref{eqU1}) comes from the derivative $%
\mbox{d}\mathbf{K}/\mbox{d}\kappa $. Thus it is natural to look for the
solution in the form 
\begin{equation}
\langle \hat{a}_{1}^{\dagger }\hat{a}_{3}\,\rangle =\frac{2\sqrt{3}}{3\pi
^{2}\kappa }\left[ A(\kappa )\mathbf{K}^{2}(\kappa )+B(\kappa )\mathbf{K}%
(\kappa )\mathbf{E}(\kappa )+C(\kappa )\mathbf{E}^{2}(\kappa )\right] ,
\label{try}
\end{equation}%
where $A(\kappa )$, $B(\kappa )$ and $C(\kappa )$ are some polynomials of $%
\kappa $. Putting the expression (\ref{try}) into equation (\ref{eqU1}) we
obtain a set of coupled equations for the coefficients of these polynomials,
which can be resolved recursively.
The equations for  other second-order moments
can be integrated in the same manner.


\newpage


\begin{figure}
\caption{
The covariance entanglement coefficient squared ${\cal Y}^2$
(thick line) and
the purity entanglement coefficient $\widetilde{\cal L}$ (thin line)
versus ``slow time'' $\tau$
for two interacting modes $\{1,1,1\}$ and $\{5,1,1\}$ in a 3D cubical cavity
($\nu=50/3$) under the condition of strict (``symmetric'') resonance and
for the initial vacuum state.
}
\label{figurYLsym}
\end{figure}

\begin{figure}
\caption{
The entropic entanglement measure $I_c$ (\ref{entr13}) versus
``slow time'' $\tau$
for two interacting modes $\{1,1,1\}$ and $\{5,1,1\}$ in a 3D cubical cavity
($\nu=50/3$) under the condition of strict (``symmetric'') resonance,
for the initial vacuum state
(thick line; $\theta_1=\theta_3=1$)
and high-temperature state (thin line; $\theta_1=3\theta_3$).
}
\label{figIsym}
\end{figure}

\begin{figure}
\caption{
The functions $\widetilde{\cal L}(\tau)$ (thin line)
and ${\cal Y}^2(\tau)$ (thick line)
for two interacting modes $\{1,1,1\}$ and $\{5,1,1\}$ in a 3D cubical cavity
($\nu=50/3$) under the condition of strict (``symmetric'') resonance and
for the high-temperature initial state with $\theta_1=140$.
}
\label{figurYLJsym}
\end{figure}

\begin{figure}
\caption{
Time dependences of different entanglement measures
under the condition of ``asymmetric resonance'' (\ref{asymm}),
for $\nu=100$ and the initial vacuum state.
Thick line:
the covariance entanglement coefficient ${\cal Y}(\tau)$ (\ref{Y}).
Thin lines from bottom to top:
the purity entanglement coefficient $\widetilde{\cal L}$ (\ref{Ltil13}),
the compact entropic entanglement measure ${\cal J}_c(\tau)$
(\ref{defJcnew}),
the function $[\widetilde{\cal L}(\tau)]^{1/2} $.
}
\label{figYJLasym}
\end{figure}

\begin{figure}
\caption{
The covariance entanglement coefficient ${\cal Y}(\tau)$ (\ref{Y})
and the purity entanglement coefficient $\widetilde{\cal L}(\tau)$
(\ref{Ltil13})
under the condition of ``asymmetric resonance'' (\ref{asymm})
with $\nu=100$,
for the initial vacuum state with $\theta_1=\theta_3=1$
(monotonous dependences)
and high-temperature state with $\theta_1=3\theta_3$
(oscillating functions). In both cases, upper curves correspond to
${\cal Y}(\tau)$ and lower curves correspond to $\widetilde{\cal L}(\tau)$.
}
\label{figY-Lasym}
\end{figure}

\begin{figure}
\caption{
The covariance entanglement coefficient ${\cal Y}_{n,m}$ (\ref{Yrsdef})
in the 1D resonance ($p=2$)  cavity versus
the compact parameter $\kappa =\tanh(2\tau)$  for
the vacuum initial state. Full curve: ${\cal Y}_{1,3}$;
dashed curve: ${\cal Y}_{3,5}$.
}
\label{figY13}
\end{figure}

\begin{figure}
\caption{
Different coefficients characterizing entanglement between the first and
third modes in the 1D resonance ($p=2$)  cavity versus ``slow time''
$\tau$ (in the insertion) and
the compact parameter $\kappa =\tanh(2\tau)$,
for the vacuum initial state.
The order of the curves in the main figure, from top to bottom:
covariance entanglement coefficient ${\cal Y}$ (\ref{Y13});
compact entropic coefficient ${\cal J}_c$ (\ref{defJcnew});
purity entanglement coefficient $\widetilde{\cal L}$ (\ref{Ltil13});
the square of the covariance entanglement coefficient ${\cal Y}^2$
(dashed curve in the insertion);
distance entanglement coefficient ${\cal Z}$ (\ref{ZGaus}).
}
\label{figYLJZ2p}
\end{figure}

\begin{figure}
\caption{
The covariance entanglement coefficient ${\cal Y}_{1,2}$ (\ref{Yrsdef})
in the 1D ``semiresonance'' ($p=1$) cavity versus the ``slow time'' $\tau $,
for the Fock (dashed curves) and squeezed vacuum (full curves) initial states
of the first mode with mean photon numbers $\nu_1=1,50,1000$.
}
\label{figY12}
\end{figure}

\begin{figure}
\caption{
The mean number of photons in the first and second modes
of the 1D ``semiresonance'' ($p=1$) cavity versus the ``slow time'' $\tau $,
for the initial Fock state $|1\rangle$.
}
\label{figmeannum}
\end{figure}

\begin{figure}
\caption{
The purity entanglement coefficient $\widetilde{\cal L}_{1,2}$
(\ref{Lp2}) (full curves)
and distance entanglement coefficient ${\cal Z}_{1,2}$ (\ref{Zp2})
(dashed curves)
versus the ``slow time'' $\tau $,
for the 1D ``semiresonance'' ($p=1$) cavity
and the initial squeezed vacuum state of the first mode
with different mean numbers of photons $\nu_1=1,50,1000$.
}
\label{figLZ1}
\end{figure}

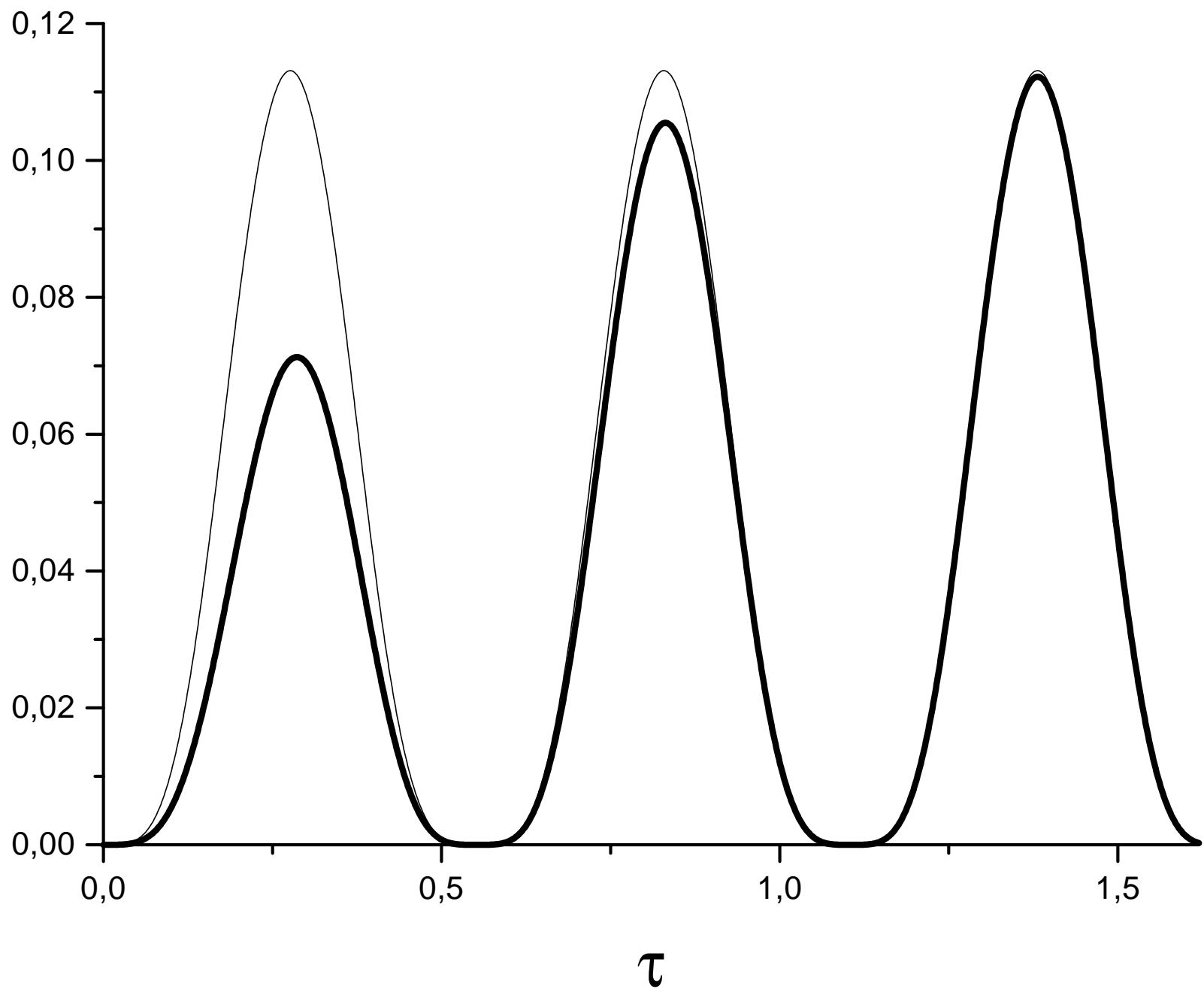
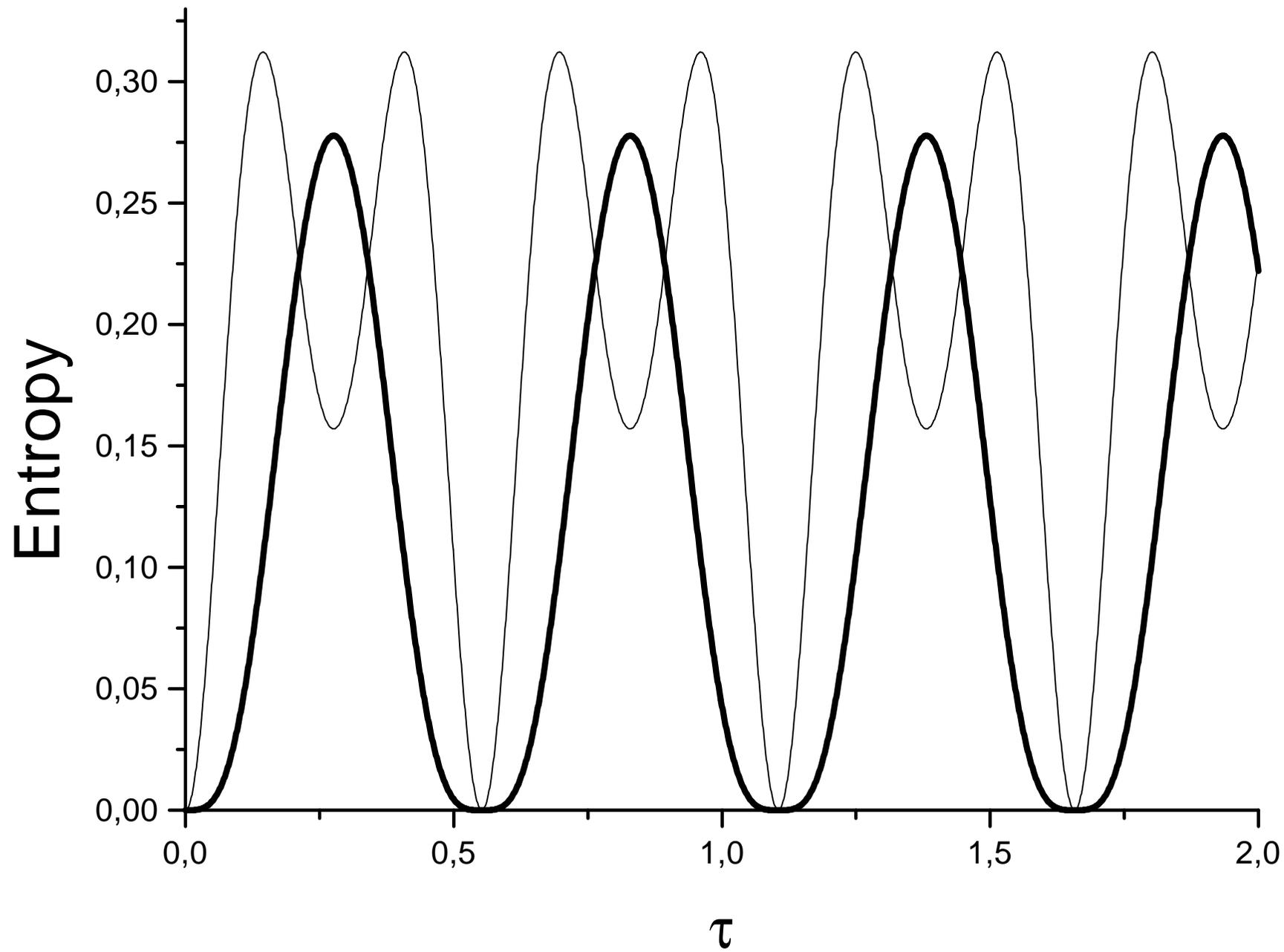
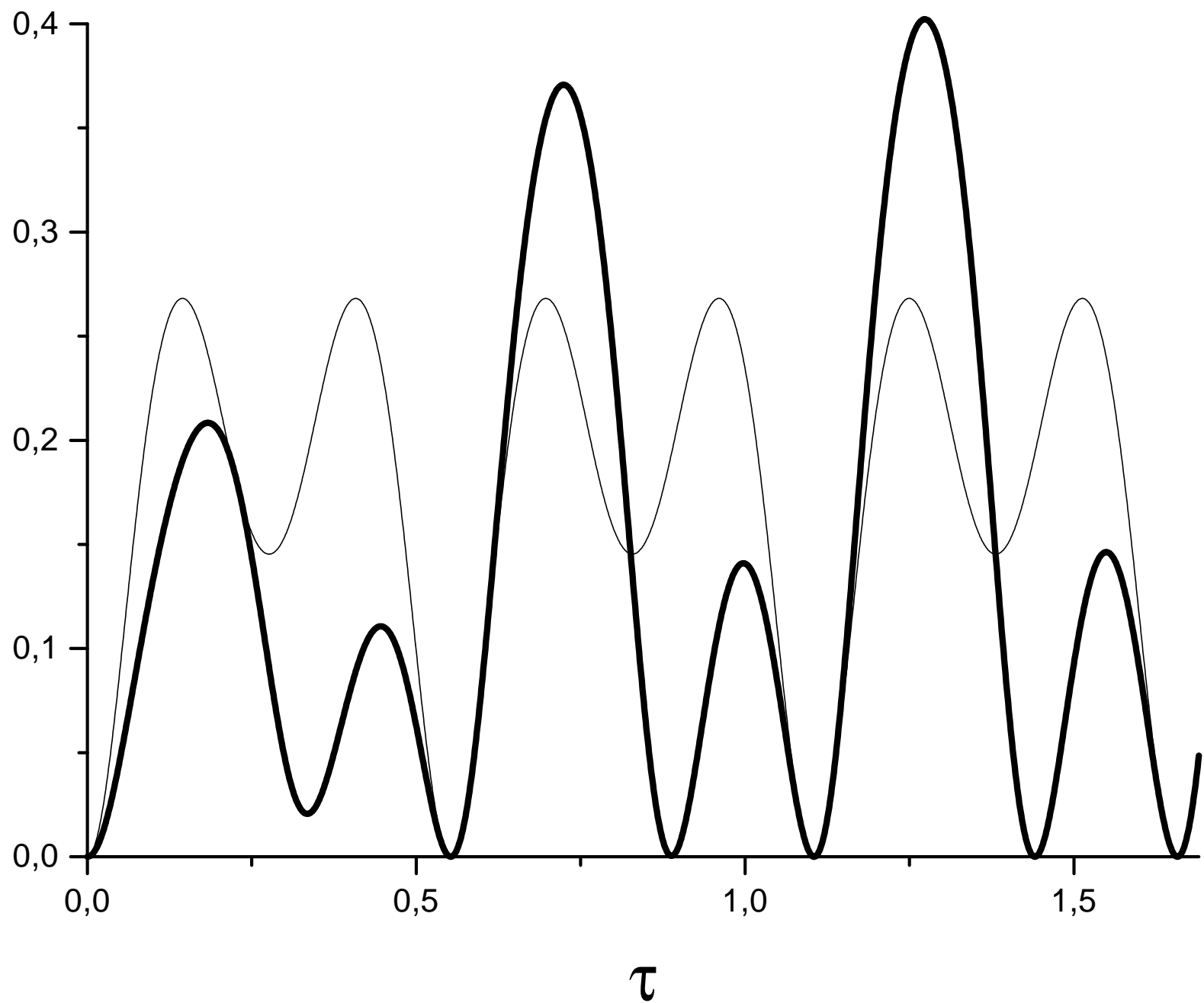
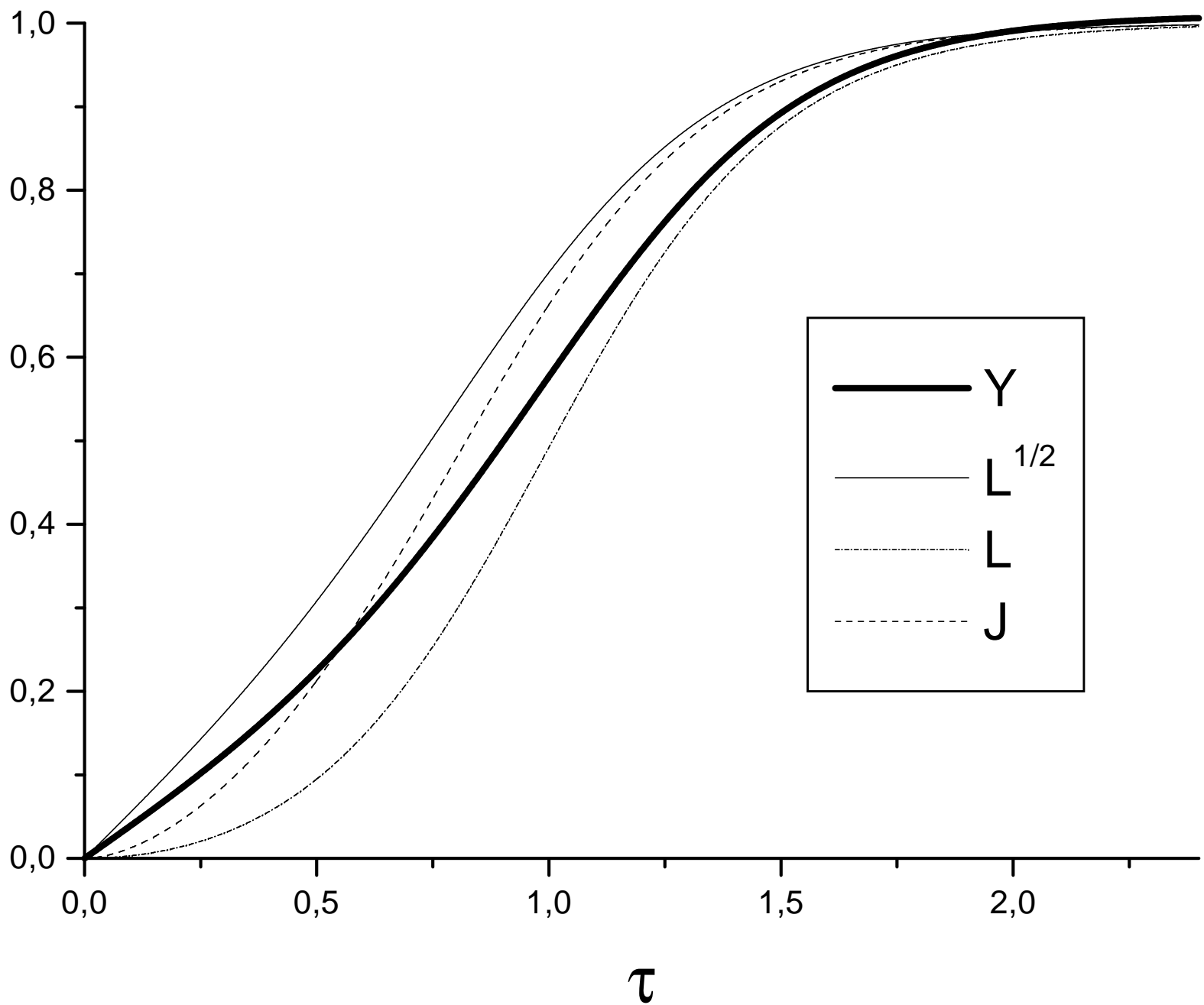
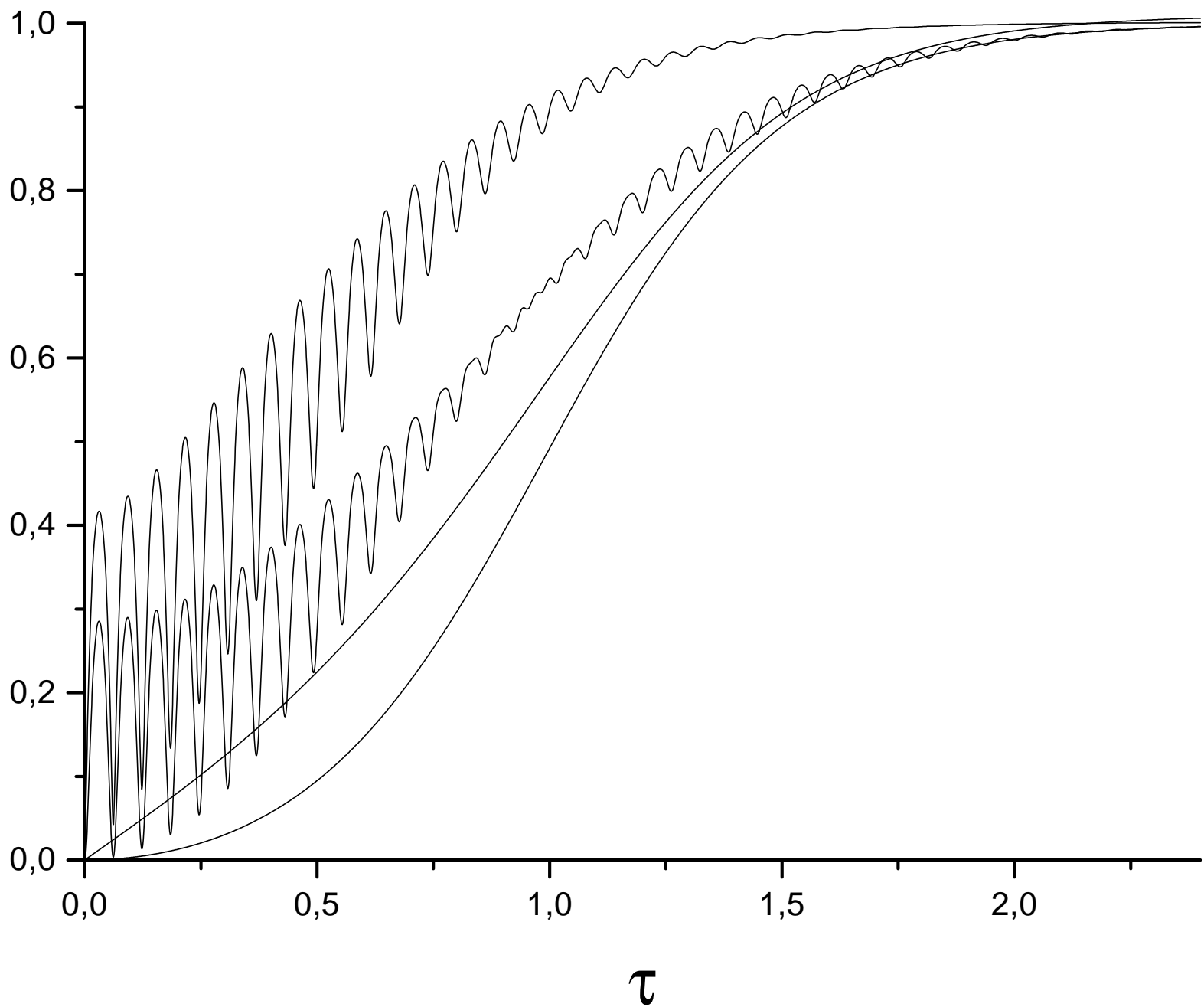
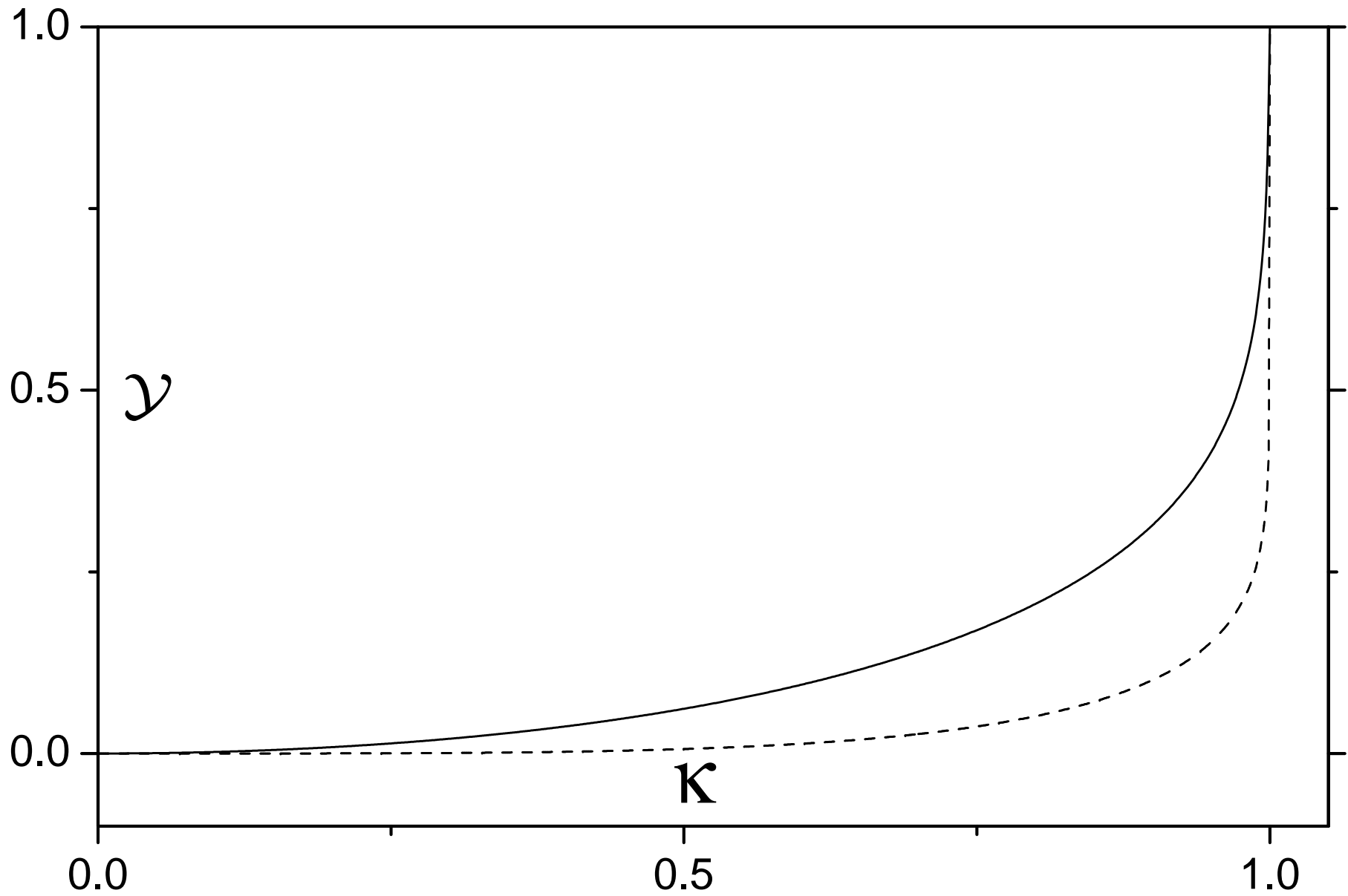
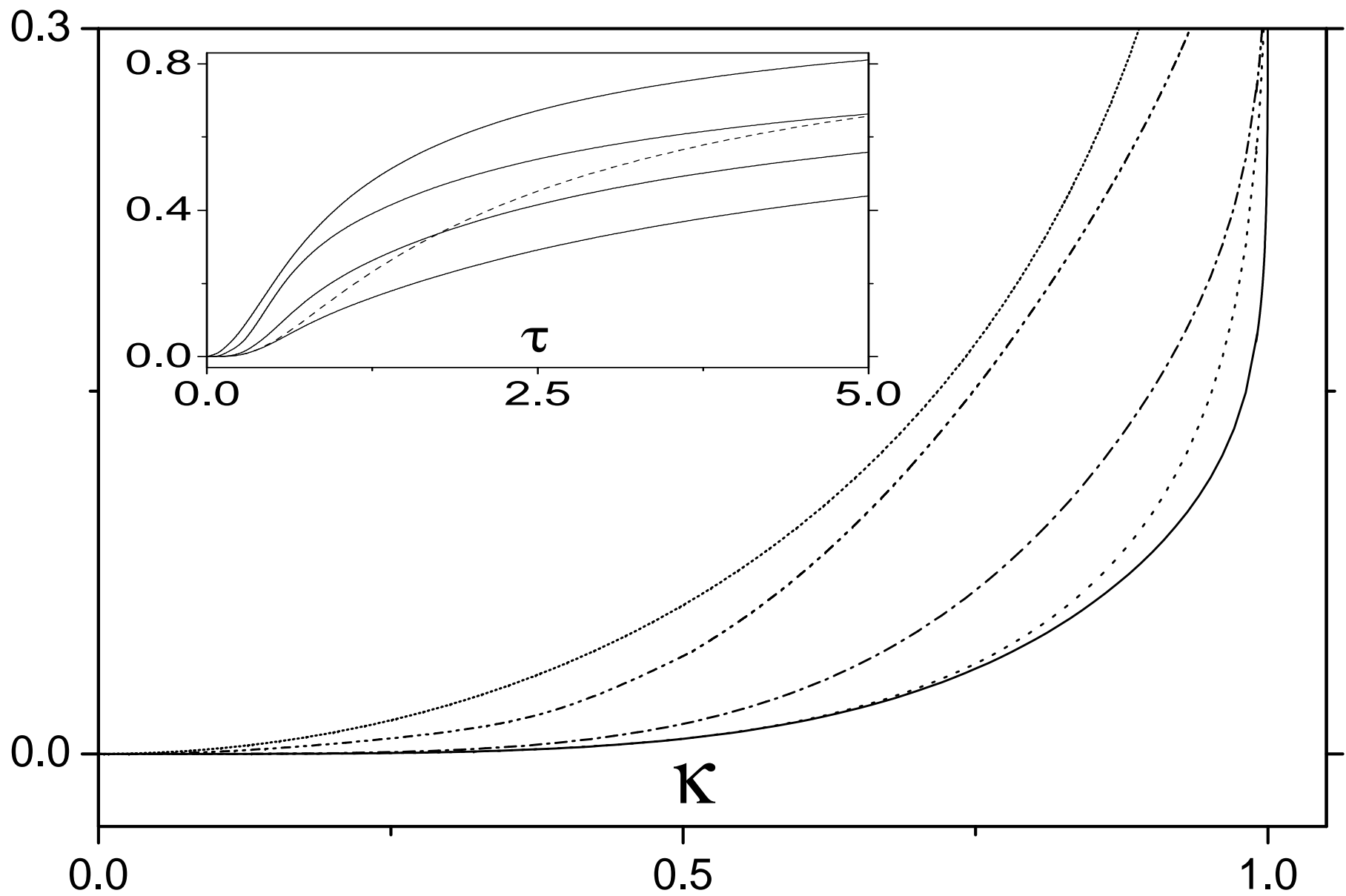
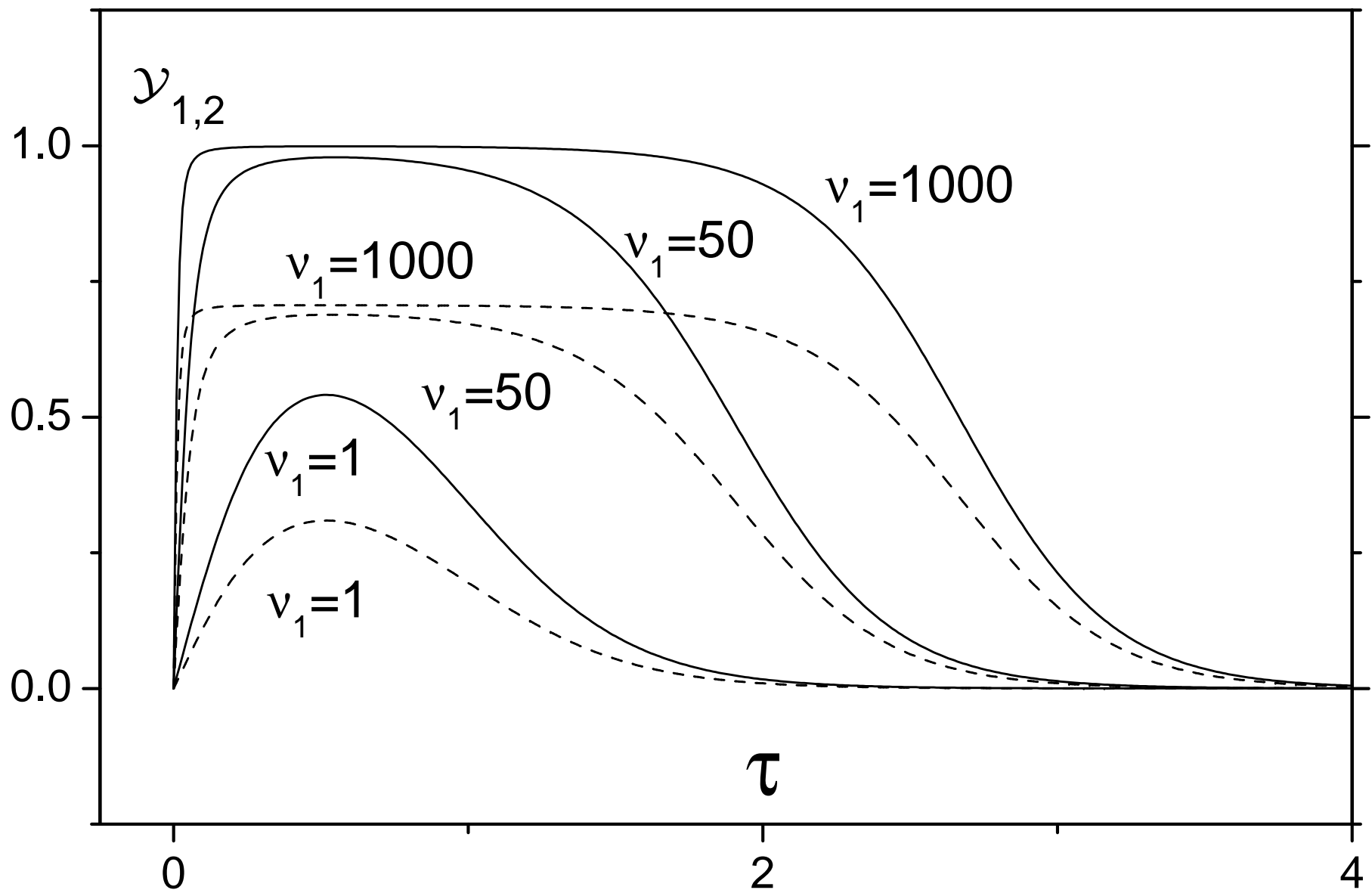
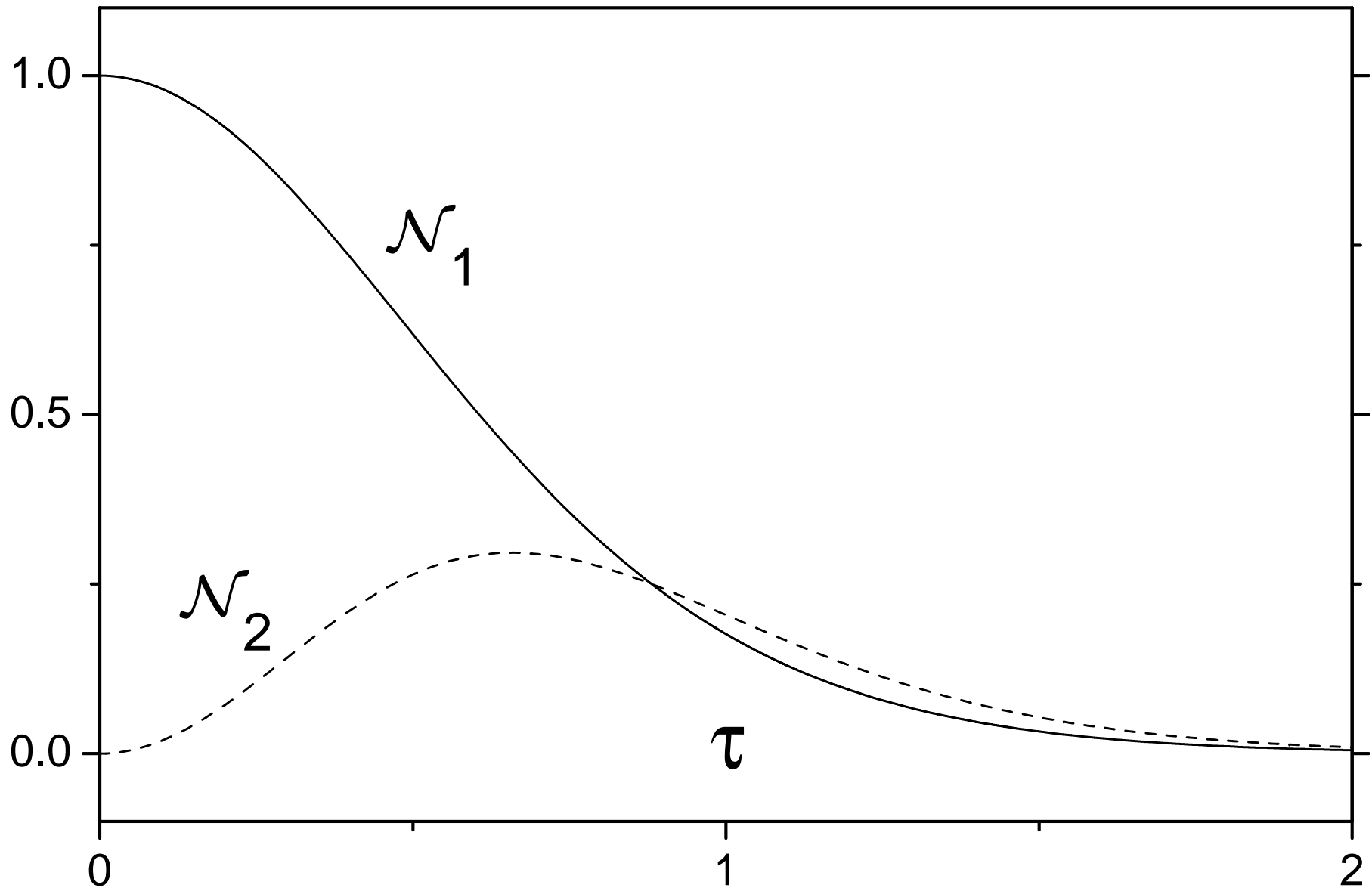
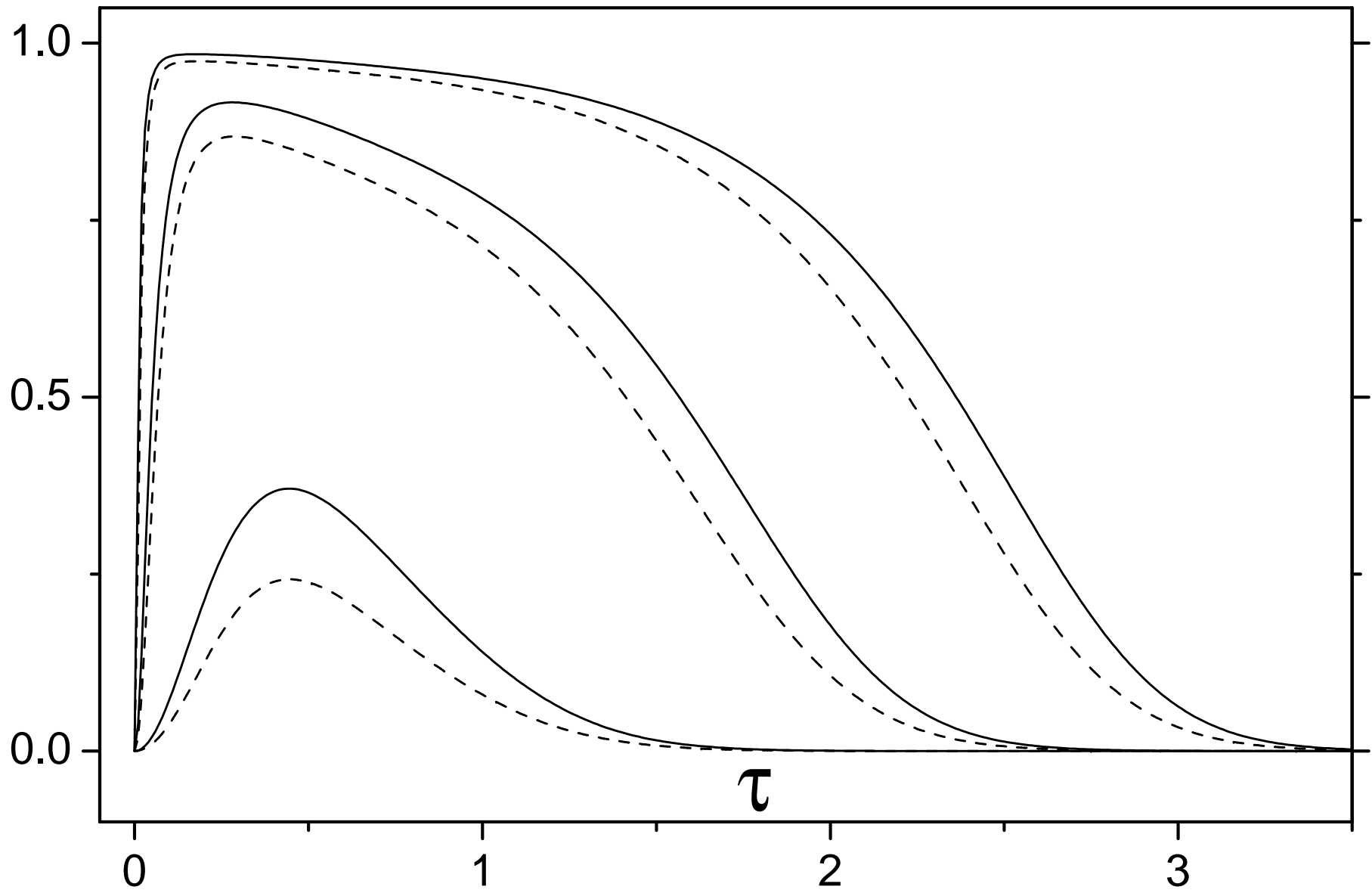
\end{document}